\documentclass[letterpaper,11pt]{article}
\usepackage{jhepmod} 

\usepackage{booktabs}
\usepackage[english]{babel}
\usepackage{amsmath,amssymb,amsbsy,amstext, amsthm, simplewick, amsfonts}
\usepackage{hyperref}
\usepackage{graphicx}
\usepackage{subcaption}

\usepackage{siunitx}
\usepackage{upgreek}
\usepackage{framed}
\usepackage{wrapfig}
\usepackage{multirow}
\usepackage{bbm}
\usepackage[svgnames,dvipsnames,x11names]{xcolor}
\usepackage[utf8x]{inputenc}
\usepackage{selinput}
\usepackage{bm}
\usepackage{float}
\usepackage{yfonts}
\usepackage{mathtools}
\usepackage{enumitem}   
\usepackage{longtable}
\usepackage{tabularx}
\usepackage{tcolorbox}
\usepackage{array}

\usepackage{tikz}
\usetikzlibrary{arrows,chains,matrix,positioning,scopes}
\usepackage{tikz-cd}
\usetikzlibrary{matrix, calc, arrows}

\allowdisplaybreaks

\def\sraise{\;\raise1.0pt\hbox{$'$}\hskip-6pt\partial}
\def\slower{\;\overline{\raise1.0pt\hbox{$'$}\hskip-6pt\partial}}

\usepackage{colortbl}

\makeatletter
\newlength{\apb@width}
\newcommand{\autoparbox}[2][c]{\settowidth{\apb@width}{#2}\parbox[#1]{\apb@width}{#2}}

\makeatother

\def\b{{\boldsymbol b}}

\def\d{{\rm d}}

\def\k{{\boldsymbol k}}

\def\n{{\boldsymbol n}}
\def\q{{\boldsymbol q}}

\def\r{{\boldsymbol r}}
\def\s{{\boldsymbol s}}

\def\u{{\boldsymbol u}}
\def\v{{\boldsymbol v}}

\def\x{{\boldsymbol x}}

\def\0{{\boldsymbol{0}}}

\def\D{{\cal D}}

\def\F{{\mathsf F}}
\def\H{{\cal H}}
\def\G{{\cal G}}
\def\K{{\cal K}}
\def\I{{\sf I}}

\def\K{{\sf K}}

\def\N{{\cal N}}
\def\M{{\cal M}}

\def\O{{\cal O}}
\def\Q{{\cal Q}}

\def\W{{\sf W}}
\def\X{{\cal X}}
\def\Y{{\cal Y}}
\def\sX{{\mathsf X}}
\def\sY{{\mathsf Y}}
\def\Z{{\cal Z}}
\def\eV{\hs\text{eV}}

\def\nn{\nonumber\\}

\def\Ddelta{\delta_{\rm D}}

\def\wt{\widetilde}

\def\fnl{f_{\rm NL}}

\def\hMpc{\, h\hs\text{Mpc}^{-1}}

\def\hs{\hskip 1pt}
\def\bhs{\hskip -0.5pt}

\def\beq{\begin{equation}}
\def\eeq{\end{equation}}
\def\be{\begin{equation}}
\def\ee{\end{equation}}

\definecolor{lightgray}{rgb}{0.9, 0.9, 0.9}

\usepackage[framemethod=default]{mdframed}
\newmdenv[skipabove=7pt,
skipbelow=7pt,
rightline=false,
leftline=false,
topline=false,
bottomline=false,
backgroundcolor=gray!10,
linecolor=gray,
innerleftmargin=5pt,
innerrightmargin=5pt,
innertopmargin=5pt,
innerbottommargin=5pt,
leftmargin=0cm,
rightmargin=0cm,
linewidth=4pt]{eBox}

\title{Precise and Accurate Cosmology with \\[5pt] CMB$\boldsymbol{\times}$LSS Power Spectra and Bispectra}

\author{Shu-Fan Chen, Hayden Lee, and Cora Dvorkin}
\affiliation{Department of Physics, Harvard University, 17 Oxford Street, Cambridge, MA 02138, USA}
\emailAdd{shufan\_chen@g.harvard.edu, hylee@g.harvard.edu, cdvorkin@g.harvard.edu}

\abstract{ 
With the advent of a new generation of cosmological experiments that will provide high-precision measurements of the cosmic microwave background (CMB) and galaxies in the large-scale structure, it is pertinent to examine the potential of performing a joint analysis of multiple cosmological probes. 
In this paper, we study the cosmological information content contained in the one-loop power spectra and tree bispectra of galaxies cross-correlated with CMB lensing. 
We use the FFTLog method to compute angular correlations in spherical harmonic space, applicable for wide angles that can be accessed by forthcoming galaxy surveys. 
We find that adding the bispectra and cross-correlations with CMB lensing offers a significant improvement in parameter constraints, including those on the total neutrino mass, $M_\nu$, and local non-Gaussianity amplitude, $\fnl$. 
In particular, our results suggest that the combination of the Vera C.~Rubin Observatory's Legacy Survey of Space and Time (LSST) and CMB-S4 will be able to achieve $\sigma(M_\nu)=42\,$meV from galaxy and CMB lensing correlations, and $\sigma(M_\nu)=12\,$meV when further combined with the CMB temperature and polarization data, without any prior on the optical depth. 
}

\begin{document}

\maketitle
\flushbottom
\newpage
\setlength{\parskip}{4pt}

\section{Introduction}

It is a remarkable fact that our seemingly complex universe can be well described by a simple parametrization of the standard cosmological model. Impressive though it may be, foundational questions regarding the nature of dark matter, the accelerated expansion rate, and the physics of the early universe remain largely unanswered. To gain further observational clues, the next generation of cosmic microwave background (CMB) experiments and large-scale structure (LSS) surveys are set to further refine measurements of cosmological parameters that characterize the geometry and late-time growth of structures in the universe.

The central observables in cosmology are correlations of matter density perturbations in the universe. As these---mostly dominated by non-luminous dark matter---are not directly observable, we infer them by measuring correlations of tracers of the underlying matter density. One important such tracer is the weak lensing of the CMB~\cite{Lewis:2006fu}, which depends on the entire trajectory of photons traveling from the last scattering surface to $z=0$, with its intensity peaking at $z\approx 2$. In particular, CMB lensing probes the same matter distribution as other LSS tracers such as galaxies, which consequently makes them highly correlated with each other. 
So far Planck has detected CMB lensing with a statistical significance of 40$\sigma$~\cite{Aghanim:2018oex}, while in the near future CMB Stage-4 (CMB-S4) is expected to further enhance this significance by an order of magnitude~\cite{Abazajian:2016yjj, Abazajian:2019eic}. This makes cross-correlations between CMB lensing and galaxies a highly appealing cosmological probe, and stimulates a joint analysis between CMB-S4 and forthcoming high-redshift ($z\gtrsim 2$) galaxy surveys such as the Vera C.~Rubin Observatory's Legacy Survey of Space and Time (LSST)~\cite{Abell:2009aa}.

One of the key physics motivations for cross-correlations of CMB lensing with galaxy clustering is to measure the total mass $M_\nu$ of the three neutrino species. Neutrino oscillation measurements have informed us that neutrinos must be massive and that their total mass must be bounded below by $60$~or~$100\,$meV for the normal and inverted hierarchies,\footnote{More specifically, the mass-squared splittings are given by $m_2^2-m_1^2\simeq 7.6\times 10^{-5}\,$eV$^2$ and $|m_3^2-m_1^2|\simeq 2.5\times 10^{-3}\,$eV$^2$. Due to the unknown sign for the latter, there can be two scenarios with $m_1<m_2<m_3$ (normal hierarchy) and $m_3<m_1<m_2$ (inverted hierarchy).} respectively~\cite{Maltoni:2004ei}. While Planck's current 95\% upper limit of $M_\nu\le 120\,$meV~\cite{Aghanim:2018eyx} is compatible with both mass ordering scenarios, the combination of CMB-S4 and LSST has strong prospects to determine the neutrino mass hierarchy over the next decade. 
Another important motivation is to improve the current constraint on the parameter $\fnl$ that characterizes the size of primordial non-Gaussianity, especially that of the local type which leaves the largest imprint on LSS observables. A realistic observational target for upcoming LSS experiments is to reach $\sigma(\fnl)\simeq 1$ for local non-Gaussianity, improving upon the state-of-the-art constraint $\fnl=-0.9\pm 5.1\,(68\%\ \text{CL})$ from Planck~\cite{Akrami:2019izv}. Achieving this will give us a better insight into the inflationary dynamics as $\fnl\simeq 1$ is the natural level of non-Gaussianity generated in many physically motivated inflationary scenarios~\cite{Alvarez:2014vva}.

Given that upcoming experiments will allow us to measure correlations of CMB lensing and galaxies with an unprecedented accuracy, it is equally important to be able to provide precise theoretical predictions for these observables. 
Being an integrated measure of the gravitational potential along the line of sight, CMB lensing is intrinsically defined on a two-dimensional celestial sphere. 
Moreover, galaxies are mapped by their redshifts and angular positions on the sky. 
The observables of interest are therefore (cross-)correlation functions of CMB lensing and galaxies in angular space.
There is, however, a well-known numerical challenge associated with computing angular observables, namely that the two-dimensional projection integrals involve highly-oscillatory Bessel functions that complicate their numerical evaluation. 
The usual workaround is to adopt the so-called Limber approximation~\cite{Limber:1954zz}, which replaces the Bessel function with a Dirac delta function located at its first peak. 
Though surprisingly accurate, this naive substitution also comes with several limitations: it is only valid for correlations over small angular separations, with sufficiently wide and overlapping tomographic bins.
These assumptions will no longer be fully justified for forthcoming galaxy surveys, which will probe wider angles with smaller redshift measurements errors. An accurate parameter inference will therefore require a means of efficiently computing angular correlations without relying on these assumptions.

In recent years there have been great advancements in the development of efficient approaches to cosmological correlation computations. A particularly noteworthy method is  FFTLog~\cite{Hamilton:1999uv} and its modern incarnations for computing Fourier-space correlators~\cite{McEwen:2016fjn, Schmittfull:2016jsw, Schmittfull:2016yqx, Simonovic:2017mhp}. The novelty of the FFTLog-based methods is to expand the linear power spectrum as a finite sum over complex power-law functions (i.e.~FFT~in~$\log k$), after which the integrals associated with each summand can be done {\it analytically}. The resulting analytic expressions can then be efficiently evaluated and summed over with pre-computed coefficients. 
It turns out that this approach greatly enhance both the computational speed and numerical stability over brute-force integration, thus opening up a new path towards precision cosmology. FFTLog has also been applied to angular correlations in \cite{Assassi:2017lea, Gebhardt:2017chz} (see also \cite{Schoneberg:2018fis, Fang:2019xat, Lee:2020ebj}). This new analytical approach allows us to circumvent the oscillatory part of the projection integrals, providing a substantial numerical advantage compared to existing methods, without being tied to the limitations of the Limber approximation.

The aim of the present work is to use FFTLog to provide a precise and accurate forecast on cosmological parameters from the combination of CMB-S4 with LSST, extending the previous works~\cite{Schmittfull:2017ffw, Yu:2018tem} in a number of directions. Specifically, the observables considered in our analysis are the angular power spectra and bispectra of CMB lensing and galaxy overdensity in spherical harmonic space, as well as their cross-spectra. For the power spectra, we also include one-loop corrections from perturbation theory. 
We will present a detailed discussion on its impact on parameter constraints as well as that of various other contributions such as the redshift space distortion, tomographic cross-spectra, and a CMB prior. Also discussed will be the biases induced by the Limber approximation in parameter estimation and its comparison with FFTLog.

\paragraph{Outline} 
The outline of the paper is as follows. In Section~\ref{sec:th}, we describe the theoretical model we employ in our analysis on the evolution of matter and galaxy densities in the presence of massive neutrinos. In Section~\ref{sec:ang}, we describe cross-correlations between galaxy and CMB lensing in harmonic space. In Section~\ref{sec:fisher}, we forecast cosmological constraints from future CMB and LSS experiments, highlighting the constraint on the neutrino mass and $\fnl$. We conclude in~Section~\ref{sec:con}. 
A number of appendices contain supplementary materials. Appendix~\ref{app:bias} describes the bias evolution model that we use. Appendix~\ref{app:loop} contains the expressions for the one-loop integrals. Appendix~\ref{app:more} collects experimental specifications and compares the forecasts between different experiments. Finally, Appendix~\ref{app:analytic} presents analytic formulas for evaluating angular power spectra.

\section{Theoretical Model for Galaxy Clustering}\label{sec:th}

We begin with a review of perturbative methods in cosmology, and describe the theoretical model that we work with in our analysis. 
We mostly describe observables in Fourier space in this section, leaving the discussion of their harmonic space counterparts to~\S\ref{sec:ang}. In~\S\ref{sec:matter}, we summarize the standard perturbative approach to matter clustering and its modifications in the presence of massive neutrinos and primordial non-Gaussianity. We then describe how matter overdensity is related to galaxy clustering in~\S\ref{sec:galaxy}. 
This section consists mostly of review materials, and experts who are interested in details of the computation may skip to~\S\ref{sec:ang}.

\subsection{Perturbation Theory with Neutrinos}\label{sec:matter}

The main constituents of matter in the universe are cold dark matter (CDM) and baryons, while neutrinos make up a small fraction of the total matter density. We therefore start with a short summary of the standard cosmological perturbation theory (SPT) with zero neutrino density~\cite{Bernardeau:2001qr}. 
In this framework, CDM is treated as an effective pressureless fluid on large scales. The evolution of its density contrast $\delta$ and its velocity divergence $\theta\equiv\nabla\cdot\v$ are then described by the continuity and Euler equations, which in Fourier space take the form
\beq\label{eom}
\begin{aligned}
	\delta' + \theta & = -[\theta\star\delta]_\alpha\, ,\\
	\theta'+\H\theta+\frac{3}{2}\H^2\Omega_m\delta & = -[\theta\star\theta]_\beta\, ,
\end{aligned}
\eeq
where a prime denotes a derivative with respect to conformal time $\tau$, $\H$ is the conformal Hubble parameter, and $\Phi_g$ is the gravitational potential that obeys the Poisson equation $\nabla^2\Phi_g = \frac{3}{2}\H^2\Omega_m\delta$.
The convolutions in the equations of motion are given by
\begin{align}
	[\theta\star\delta]_\alpha(\k) &\equiv \int_\q \alpha(\k,\q)\theta(\q)\delta(\k-\q)\, , \quad \alpha(\k,\q) \equiv \frac{\k\cdot\q}{q^2}\, ,\\
	[\theta\star\theta]_\beta(\k) &\equiv \int_\q \beta(\k,\q,\k-\q)\theta(\q)\delta(\k-\q)\, , \quad\beta(\k,\q_1,\q_2)\equiv \frac{k^2(\q_1\cdot\q_2)}{2q_1^2q_2^2}\, ,
\end{align}
where $\int_{\q_1,\cdots,\q_n}\equiv\int\frac{\d^3q_1}{(2\pi)^3}\cdots\frac{\d^3q_n}{(2\pi)^3}$. On large, quasi-linear scales, the equations \eqref{eom} can be solved order by order in perturbation theory.\footnote{The nonlinear scale can be identified as the scale $k_{\rm NL}$ at which the dimensionless matter power spectrum becomes unity. For $k\gtrsim k_{\rm NL}$, the perturbative expansion ceases to converge and one has to resort to either numerical simulations or empirical models.} 
While obtaining the solutions in a generic $\Lambda$CDM cosmology is rather complicated, a great simplification can be made by assuming that the growth rate of perturbations is the same as in the matter-only, Einstein-de Sitter (EdS) universe, which is known to be valid to a few percent accuracy~\cite{Blas:2014hya, Donath:2020abv}. Under this assumption, the temporal and spatial dependences of the higher-order solutions factorize as
\begin{align}
	\delta(z,\k) = \sum_{n=1}^\infty D_+(z)\delta_n(\k)\, , \label{EdSsol}
\end{align}
and similarly for $\theta$, where the linear growth function takes the form $D_+(z)=(1+z)^{-1}$ in the EdS universe, normalized to unity at $z=0$. The $n$-th order solution is given in terms of the linear solution $\bar\delta(\k)\equiv\delta_1({z=0},\k)$ by
\begin{align}
	\delta_n (\k) &= (2\pi)^3\int_{\q_1,\cdots,\q_n}\Ddelta(\k-\q_{1\cdots n}) F_n(\q_1,\cdots,\q_n) \bar\delta(\q_1)\cdots \bar\delta(\q_n)\, ,\label{deltan}
\end{align}
where $\q_{1\cdots n}\equiv \q_1+\cdots+ \q_n$ and the kernels $F_n$ are rational functions of the inner products $\q_i\cdot\q_j$ that can be computed iteratively~\cite{Fry:1983cj, Goroff:1986ep, Jain:1993jh}.

A few modifications to the SPT are necessary to incorporate free-streaming massive neutrinos. The total matter density contrast is now decomposed into two components as
\begin{align}
	\delta = (1-f_\nu)\delta_{cb} + f_\nu\hs\delta_\nu\, ,
\end{align}
where $\delta_{cb}$ and $\delta_{\nu}$ are the density contrasts for CDM+baryons and neutrinos, respectively, and
\begin{align}
	f_\nu\equiv \frac{\Omega_\nu}{\Omega_m} \approx \frac{1}{\Omega_{m0}h^2}\frac{M_\nu}{93.14\hs\text{eV}}
\end{align}
denotes the fractional density of neutrinos, with $M_\nu$ being the total mass of the three neutrino species and $\Omega_{m0}$ the matter density today. Since the current constraint on the neutrino mass~\cite{Aghanim:2018eyx}
\begin{align}
	0.06\eV < M_\nu < 0.12 \eV
\end{align}
implies $f_\nu=O( 10^{-2})$, we can restrict ourselves to neutrino perturbations to linear order (see~\cite{Dvorkin:2019jgs} for a review of neutrino mass constraints over the next decade). 
The total matter power spectrum in the presence of neutrinos is then given by
\begin{align}
	P^{mm}(z,z',k)&= (1-2f_\nu)P^{{cb},{cb}}(z,z',k)+2f_\nu P^{{cb},\nu}(z,z',k)+O(f_\nu^2)\, ,
\end{align}
where $P^{cb,cb}\equiv\langle \delta_{cb}\delta_{cb}\rangle'$ and $P^{cb,\nu}\equiv\langle \delta_{cb}\delta_{\nu}\rangle'$ denote the power spectra of the different components. The presence of massive neutrinos affect the matter power spectrum through the correction $f_\nu$ as well as the modification of the growth function. This is because massive neutrinos induce a scale dependence in the evolution of matter fluctuations due to the fact that neutrinos become non-clustering below the free-streaming scale\footnote{The free-streaming of neutrinos does not only modify the growth of matter density but also affects its bias relation to galaxy distributions. This is captured by a scale-dependent correction to the linear galaxy bias~\cite{LoVerde:2014pxa,Vagnozzi:2018pwo,Munoz:2018ajr,Xu:2020fyg}.}~\cite{Lesgourgues:2006nd} 
\begin{align}\label{kfs}
	k_{\rm fs} \approx 0.023\left(\frac{M_\nu}{0.1\eV}\right)\left(\frac{2}{1+z}\frac{\Omega_{m0}}{0.23}\right)^{\frac{1}{2}}\hMpc \, .
\end{align}
An important consequence of this is that the growth function can no longer be simply factored out from the momentum dependence as in \eqref{EdSsol}, requiring a modification to the standard perturbative calculations that rely on the EdS approximation. A systematic treatment of massive neutrinos in perturbation theory using time-dependent Greens functions is given in~\cite{Senatore:2017hyk}.
We will, however, work with a simple ansatz and assume that the linear power spectrum factorizes as
\begin{align}
	P_{11}(z,z',k) \equiv \langle\delta_1(z,\k)\delta_1(z',-\k)\rangle'= D_+(z,k)D_+(z',k)P_{11}(k)\, ,\label{P11}
\end{align}
where $P_{11}(k)$ is the linear power spectrum at ${z=z'=0}$ and a prime on $\langle\hs\cdots\rangle$ means that the momentum-conserving delta function has been factored out. 
As was demonstrated in~\cite{Hu:1997vi, Takada:2005si}, this ansatz provides a good fit to the numerical result with a few percent accuracy.

\subsection{Galaxy Bias}\label{sec:galaxy}

We do not observe the distribution of matter directly but instead that of its tracers in the large-scale structure of the universe, e.g.~galaxies. The relation between the two density distributions can be modeled in a perturbative framework of galaxy biasing (see~\cite{Desjacques:2016bnm} for a review). In this subsection, we describe relevant aspects of galaxy clustering and its relation to matter density.

It is well known that the SPT framework breaks down beyond leading order in perturbation theory due to uncontrolled ultraviolet divergences. A consistent perturbative framework is instead provided by the effective field theory of large-scale structure (EFTofLSS)~\cite{Baumann:2010tm, Carrasco:2012cv}, which extends the SPT by systematically dealing with unphysical divergences through renormalization. 
The galaxy density field is then represented as a local functional, not of the matter density, but of the potentials $\Phi\in\{\Phi_g\equiv \nabla^{-2}\delta_{cb},\Phi_v\equiv \nabla^{-2}\theta_{cb}\}$ and their derivatives.
We have
\begin{align}
	\delta_g = \sum_\Q b_\Q\hs \Q(\Phi)\, ,\label{eq:bias1}
\end{align}
with the (renormalized) bias parameter $b_\Q$ characterizing the size of the contribution from each (renormalized) operator $\Q$,\footnote{We denote bias operators by the symbol $\Q$, and reserve the symbol $\O\in\{\delta_g,\kappa\}$ for external operators in angular correlation functions.} which can in general be both redshift and scale dependent. The redshift dependence, in particular, can be measured from simulations or directly from observations.\footnote{In practice, simulations measure halo biases, as it is difficult to simulate galaxy formation on large scales. The halo distribution can then be related to the galaxy distribution through modeling of the halo occupation distribution~\cite{Berlind:2001xk}.}
For Gaussian initial conditions, the operators appearing in the bias expansion must be consistent with the equivalence principle. This implies that potentials must appear with at least two spatial derivatives, as a function of the tidal tensor $\partial_i\partial_j\Phi$. 

At one loop, operators up to third order in $\delta_{cb}$ contribute to the galaxy power spectrum~\cite{McDonald:2006mx, Chan:2012jj, Assassi:2014fva, Senatore:2014eva}. 
We use the following basis of operators for representing the galaxy bias to third order in perturbation theory:\footnote{To avoid clutter, we suppress the subscript `$cb$' in the bias coefficients.}
\begin{equation}
	\delta_g = b_\delta \delta_{cb} + b_{\delta^2} \delta^2_{cb} + b_{\G_2} \G_2[\Phi_g]+b_{\delta^3} \delta_{cb}^3 + b_{\G_2\delta} \G_2[\Phi_g]\delta_{cb} +b_{\G_3} \G_3[\Phi_g] + b_{\Gamma_3}\Gamma_3 + O(\Phi^4)\, ,
\end{equation}
where $\G_i$ are the Galileon operators defined by
\begin{align}
	\G_2[\Phi] &\equiv (\partial_i\partial_j\Phi)^2-(\partial^2\Phi)^2 \, ,\\[3pt]
	\G_3[\Phi] &\equiv \frac{3}{2}(\partial_i\partial_j\Phi)^2\partial^2\Phi-(\partial_i\partial_j\Phi)(\partial_j\partial_k\Phi)(\partial_k\partial_i\Phi)-\frac{1}{2}(\partial^2\Phi)^3\, ,
\end{align}
and $\Gamma_3\equiv \G_2[\Phi_g]-\G_2[\Phi_v]$ is an operator that characterizes the velocity tidal effects.
The bare operators give rise to divergent one-loop integrals and therefore need to be renormalized. Some operators that give purely divergent contributions get fully absorbed by counterterms, and the only renormalized operators that give non-vanishing contributions to the one-loop power spectrum are $\delta_{cb},\delta_{cb}^2,\G_2,\Gamma_3$.
Moreover, many terms give degenerate contributions, allowing us to consider a reduced set of loop integrals.
For Gaussian initial conditions, the galaxy-matter and galaxy-galaxy power spectra at one loop are given by~\cite{Assassi:2014fva}
\begin{align}
	P^{gm} &= b_{\delta}\big(P^{{cb},m}_{11}+P_{13}^{{cb},m}+P_{22}^{cb,m}\big)
	+b_{\delta^{2}}{\cal I}_{\delta^{2}}^{{cb},m} + b_{\G_{2}}{\cal I}_{\G_{2}}^{{cb},m} + (b_{\G_{2}}+\tfrac{2}{5}b_{\Gamma_{3}}){\cal F}_{\G_{2}}^{{cb},m}\, ,\label{Pgm}\\[3pt]
	P^{gg} &= b_\delta^2 \big(P^{cb,cb}_{11}+P_{13}^{cb,cb}+P_{22}^{cb,cb}\big)
+2b_{\partial^2\delta} (k/k_*)^2 P^{cb,cb}_{11}	+2b_\delta b_{\delta^2} {\cal I}_{\delta^2}^{cb,cb} +2b_\delta b_{\G_2}{\cal I}_{\G_2}^{cb,cb} \nn[3pt]
	&+ (2b_\delta b_{\G_2}+\tfrac{4}{5}b_\delta b_{\Gamma_3}){\cal F}_{\G_2}^{cb,cb}+b_{\delta^2}^{2}{\cal I}_{\delta^2\delta^2}^{cb,cb}+b_{\G_2}^2{\cal I}_{\G_2\G_2}^{cb,cb}+2b_{\delta^2}b_{\G_2}{\cal I}_{\delta^2\G_2}^{cb,cb}\, ,\label{Pgg}
\end{align}
where $b_{\partial^2\delta}$ denotes the contribution from a higher-derivative operator and $k_*$ is a renormalization scale. For clarity, we have suppressed all arguments in the above expressions, and we present the redshift-dependent integral representations of $P_{13}$, $P_{22}$, ${\cal F}_{\O}$, ${\cal I}_{\O}$, ${\cal I}_{\O\O'}$ in Appendix~\ref{app:loop}. Note that $b_{\Gamma_3}$ does not induce a new shape dependence at one loop. Moreover, it has the same redshift dependence as $b_{\G_2}$ in certain bias models, making the two biases almost degenerate with each other. For this reason, it is often preferred to fix the value of (or impose a sharp prior on) $b_{\Gamma_3}$ in data analysis; see e.g.~\cite{Chudaykin:2019ock}.

We also consider bispectra in our analysis. For simplicity, we restrict ourselves to bispectra at tree level, in which case we take the bias expansion up to second order in $\delta_{cb}$. The gravity-induced galaxy and matter bispectra at tree level are given by\footnote{See \cite{deBelsunce:2018xtd} for an exact treatment for including massive neutrinos for the tree matter bispectrum.}
\begin{align}
	B^{mmm}&= 2D_{1,2}D_{1,3}D_{2,2}D_{3,3} P_2^{mm}P_3^{mm}F_2(\k_2,\k_3)+\text{2 perms}\, ,\label{Bmmm}\\[3pt]
	B^{mmg} &=2 D_{1,2}D_{1,3}D_{2,2}D_{3,3} P^{mm}_2P_3^{cb,m}\big(b_{\delta,3}F_{2}(\k_2,\k_3)\big)+(1\leftrightarrow2)\nn[3pt]
	&\quad+2 D_{1,1}D_{2,2}D_{3,1}D_{3,2} P_1^{cb,m}P_2^{cb,m}\big(b_{\delta^2,3} +b_{\G_2,3}L_2(\k_2,\k_3)+b_{\delta,3}F_{2}(\k_2,\k_3)\big)\, ,\\[3pt]
	B^{mgg} &= 2b_{\delta,2} D_{1,1}D_{2,2}D_{3,1}D_{3,2} P_1^{cb,m}P_2^{cb,cb}\big(b_{\delta^2,3} +b_{\G_2,3}L_2(\k_1,\k_2)+b_{\delta,3}F_{2}(\k_1,\k_2)\big) + (2\leftrightarrow 3)\nn[3pt]
	&\quad+2b_{\delta,2}b_{\delta,3} D_{1,2}D_{1,3}D_{2,2}D_{3,3} P_2^{cb,m}P_3^{cb,m}F_{2}(\k_2,\k_3)\, ,\\[3pt]
	 B^{ggg}
	 &= 2b_{\delta,1}b_{\delta,2} D_{1,1}D_{2,2}D_{3,1}D_{3,2} P_1^{cb,cb}P_2^{cb,cb}\big(b_{\delta^2,3}+b_{\G_2,3} L_2(\k_1,\k_2)+b_{\delta,3} F_2(\k_1,\k_2)\big)\nn[3pt]
	 &\quad +\text{2 perms}\label{Bggg}\, ,
\end{align}
where to avoid clutter we have suppressed some arguments and introduced the shorthand notation $b_{\O,a}\equiv b_\O(z_a)$, $D_{a,b}\equiv D_+(z_a,k_b)$, $P_a^{\O\O'}\equiv P_{11}^{\O\O'}(k_a)$, and 
${L_2(\k,\q) \equiv (\hat\k\cdot\hat\q)^2-1}$. 

\subsubsection{Non-Gaussian Initial Conditions}

We have so far described correlations that arise due to matter clustering at late times for initially Gaussian density perturbations. If the initial statistics are instead non-Gaussian, then there are additional contributions. This can be seen from the relation between the linear matter density and the primordial potential $\phi$:
\begin{align}
	\delta^{(1)}(z,\k) =  \frac{2}{3}\frac{k^2T(z,k)}{\Omega_{m0} H_0^2}\phi(\k)\equiv M(z,k)\phi(\k)\, ,
\end{align}
where $T$ is the transfer function with normalization $T(z=0,k\to 0)=1$, with $\phi$ also being related to the potential that appeared in the bias expansion in \eqref{eq:bias1} by $\Phi(z,\k) = -\frac{2 T(z,k)}{3\Omega_m H_0^2}\phi(\k)$. The bispectrum of $\phi$ then induces a nonzero matter bispectrum at tree level as
\begin{align}
	B^{mmm}_{\rm PNG}(\{z_i,k_i\}) = M(z_1,k_1)M(z_2,k_2)M(z_3,k_3)B^{\phi\phi\phi}(k_1,k_2,k_3)\, .
\end{align}
Since we are interested in the regime of weak non-Gaussianity, let us assume that the primordial potential is quadratic in a Gaussian field $\phi_g$ and write~\cite{Schmidt:2010gw, Scoccimarro:2011pz, Assassi:2015fma}
\begin{align}
	\phi(\k) = \phi_g(\k) +\fnl\int_\q K_{\rm NL}(\q,\k-\q)\big[\phi_g(\q)\phi_g(\k-\q)-\langle\phi_g(\q)\phi_g(\k-\q)\rangle\big]\, ,
\end{align}
with the nonlinear kernel $K_{\rm NL}$ parametrizing the shape of the bispectrum. This assumption neglects terms higher order in $\phi_g$ that contribute to the three-point function at loop level or higher-point functions at tree level. Note that this reduces to the familiar expression for primordial non-Gaussianity of the local type $\phi(\x)=\phi_g(\x) + \fnl(\phi_g^2(\x)-\langle\phi_g^2\rangle)$ in position space when $K_{\rm NL}=1$. The above ansatz yields the bispectrum
\begin{align}
	B^{\phi\phi\phi}(k_1,k_2,k_3)=2\fnl K_{\rm NL}(\k_1,\k_2)P^{\phi\phi}(k_1)P^{\phi\phi} (k_2) + \text{2 perms}\, ,\label{Bphi}
\end{align}
with $\frac{9}{25}\frac{k^3}{2\pi^2}P^{\phi\phi}(k)=A_s(k/k_\star)^{n_s-1}$. The nonlinear kernel may be expanded around the squeezed limit, where the wavenumber of a long-wavelength mode, $k_L$, is much smaller than the two other short-wavelength modes, $k_S\gg k_L$, as
\begin{align}
	K_{\rm NL}(\k_L,\k_S) = \sum_{n,J=0}^\infty a_{n,J}\left(\frac{k_L}{k_S}\right)^{\Delta+n}P_{J}(\hat\k_L\cdot\hat\k_S)\, ,
\end{align}
with $\Delta$ parametrizing the leading power-law scaling in the squeezed limit. In the context of inflationary model building, $\Delta$ may count the number of derivatives of the self-interaction of the curvature perturbation or reflect the mass $m$ of the particle that contributes to the bispectrum through particle production. 
This basis naturally captures the inflationary bispectra that arise from the exchange of spin-$J$ particles~\cite{Arkani-Hamed:2015bza, Lee:2016vti, Arkani-Hamed:2018kmz}.\footnote{See \cite{MoradinezhadDizgah:2018ssw} for forecasts on the galaxy bispectrum of massive spinning particles with upcoming surveys.}
The long-wavelength limit of the primordial bispectrum then leads to the scale-dependent correction $\Delta b_\delta$ to the linear bias $b_\delta$; e.g.~for $J=0$, we have
\begin{align}\label{scaledependentbias}
	\Delta b_\delta(z,k) = \frac{2\fnl(b_\delta(z)-1)\delta_c}{M(k,z)}\left(\frac{k}{q_*}\right)^\Delta +\cdots\, ,
\end{align}
where $\delta_c\approx 1.686$ is the critical overdensity, $q_*$ is some reference scale, and we have only kept the leading correction in the limit $k\to 0$. We see that this goes as $\Delta b_\delta\sim 1/k^{2-\Delta}$ as $k\to 0$, and reproduces the well-known scale-dependent bias of \cite{Dalal:2007cu} for local non-Gaussianity with $\Delta=0$. While the scaling can lie anywhere within the interval $\Delta\in[0,2]$ for conventional early-universe scenarios, we focus our attention to the case $\Delta=0$ in this work, which is the shape that can be best constrained through this effect.\footnote{See~\cite{Gleyzes:2016tdh, MoradinezhadDizgah:2017szk} for forecasts on non-Gaussianity beyond the local type from the scale-dependent bias.}

\subsection{Redshift Space Distortion}

An important contribution to galaxy clustering is the redshift space distortion (RSD).
This arises due to peculiar velocities of galaxies, which lead to an anisotropic correction to galaxy statistics on top of the Hubble flow. This presents both a challenge and an opportunity: while extra complications must be faced to deal with the RSD, it also provides an additional means of constraining the growth of structure. 

At leading order in perturbation theory, the RSD is captured by the Kaiser formula~\cite{Kaiser:1987qv}
\begin{align}
	\delta_g(z,\k) = \delta_{cb}(z,\k)(b_\delta+f_+(z,k)\mu^2)\, ,\label{RSD}
\end{align}
where $\mu\equiv \hat\k\cdot\hat\n$ is defined to be the angle between $\hat\k$ and the line-of-sight direction $\hat\n$, and $f_+\equiv \d\log D_+/\d\log a$ is the logarithmic growth function. The RSD induces non-vanishing multipole moments of the power spectrum up to quartic order in $\mu$, and allows for a measurement of the growth rate $f_+$. 
In particular, since massive neutrinos suppress the growth of structure at small scales, they also suppress the RSD contribution at small scales and renders $f_+$ scale dependent. 
An accurate measurement of the RSD can therefore provide a more accurate constraint on the total neutrino mass~\cite{Hernandez:2016xci, Boyle:2017lzt}. 

The linear RSD also introduces the following counterterms in the galaxy power spectrum at one loop:
\begin{align}
	P^{gg}(k)\, \supset\, 2f_+(z,k)(d_1 \mu^2+d_2 \mu^4)(k^2/k_*^2)P_{11}(k)\, ,
\end{align}
with $d_1, d_2$ unfixed parameters, as well as modifying the one-loop integrals. The RSD in general gives a subleading contribution to the power spectrum on large scales for typical galaxy window functions with wide tomographic bins. The additional RSD terms at one loop then only give small corrections, and we therefore restrict ourselves to the tree-level RSD in our analysis. We also neglect other types of relativistic effects that become relevant at very large angular scales, such as the (integrated) Sachs-Wolfe effect and the Doppler shift~\cite{Yoo:2009au, Bonvin:2011bg, Challinor:2011bk}, which give subdominant contributions compared to the RSD in the multipole ranges that we consider in this work.

\section{Harmonic Space Analysis}\label{sec:ang}

While galaxies have a three-dimensional distribution, CMB lensing depends only on the line-of-sight direction and is an intrinsically two dimensional observable. Analyzing the combined statistical properties of galaxies and CMB lensing therefore requires taking their cross-correlations in two-dimensional (spherical) harmonic space. 

We begin by describing projected observables in harmonic space in \S\ref{sec:obs}. We then introduce analytical tools for evaluating the angular projection integrals in \S\ref{sec:gen}, and describe the computation of the angular power spectrum and bispectrum in \S\ref{sec:angcorr}. 
New results are presented in \S\ref{sec:poly}, where we apply the polynomial approximation to scale-dependent integral kernels and obtain analytical formulas for approximating the angular power spectrum.

\subsection{Projected Observables}\label{sec:obs}

A projected operator $\O$ along the line-of-sight direction $\hat\n$ takes the form
\begin{align}
	\O(\hat\n) &= \int_0^\infty \d\chi\, W_\O(\chi)\O(\chi,\chi\hat\n) = \sum_{\ell m}\O_{\ell m}Y_{\ell m}(\hat\n)\, ,\label{Olm}
\end{align}
where $\sum_{\ell m}\equiv \sum_{\ell=0}^\infty\sum_{-\ell\le m\le \ell}$, $\chi$ is comoving distance and $W_\O$ is a window function for $\O$, and we have expanded the operator in terms of the spherical harmonics $Y_{\ell m}$. Abusing the notation, we will sometimes distinguish the $z$ and $\chi$ dependence of a function only by its arguments, e.g.~$\O(\chi,\chi\hat\n)=\O(z,\chi\hat\n)$. The orthonormality of the spherical harmonics implies that the harmonic coefficient $\O_{\ell m}$ can be obtained as
\begin{align}
	\O_{\ell m} = 4\pi i^\ell\int_0^\infty \d\chi\, W_\O(\chi)\int_\k \, j_\ell(k\chi)Y_{\ell m}^*(\hat\k)\widetilde\O(z,\k)\, ,
\end{align}
where we have used the plane-wave expansion $	e^{i\k\cdot\r} = 4\pi\sum_{\ell m} i^\ell j_\ell(kr)Y_{\ell m}^*(\hat\k)Y_{\ell m}(\hat\r)$ to relate the harmonic coefficient to the Fourier-space operator $\widetilde\O$. In this work, we will be interested in two types of projected operators: the CMB lensing convergence $\kappa$ and tomographic galaxy overdensity $\delta_g$.

CMB lensing is an integrated measure of the gravitational potential up to the last scattering surface. It therefore probes the matter distribution in the late universe and is complementary to other tracers of the large-scale structure. 
More precisely, the CMB lensing potential $\psi$ is defined as an effective integrated potential along the line of sight, given by~\cite{Bartelmann:1999yn, Lewis:2006fu}
\begin{align}
	\psi(\hat\n) = -2\int_0^\infty\d\chi\hs \frac{\chi_*-\chi}{\chi\chi_*}\,\Theta(\chi_*-\chi)\phi(\chi,\chi\hat\n)\, ,
\end{align}
where $\chi_*$ is the comoving distance to the last scattering surface and $\phi$ is the three-dimensional gravitational potential related to the matter overdensity by the Poisson equation $\nabla^2\phi = -\frac{3}{2}\Omega_{m0} H_0^2(1+z)\delta$. 
For weak lensing, the deflection angle is given by the two-dimensional gradient of the lensing potential, $\nabla_{\hat\n}\psi$. The CMB lensing convergence, defined as $\kappa(\hat\n)\equiv-\tfrac{1}{2}\nabla_{\hat\n}^{2}\psi(\hat\n)$, then captures the intensity magnification due to weak lensing. This quantity is directly related to the total matter overdensity as
\begin{align}
	\kappa(\hat{\n}) &= \int_{0}^{\infty}\d\chi\, W_{\kappa}(\chi)\delta(\chi,\chi\hat{\n})\, ,\\[3pt]
		W_{\kappa}(\chi) &\equiv \frac{3}{2}\Omega_{m0}H_{0}^{2}\frac{1+z(\chi)}{H(\chi)}\frac{\chi(\chi_*-\chi)}{\chi_*}\Theta(\chi_*-\chi)\, .
\end{align}
Although the lensing window function $W_{\kappa}$ has a broad kernel that extends up to $z(\chi_*)\approx 1100$, it peaks around $z\approx 2$ and is highly correlated with the galaxy distribution probed by high-redshift surveys~\cite{Schmittfull:2017ffw}.

In redshift surveys, galaxies are mapped by their angular positions on the sky and their redshifts. Due to the imprecise nature of redshift measurements, observed galaxy samples are split into tomographic bins of finite widths. The galaxy overdensity for the $i$-th tomographic bin, denoted $\delta^{(i)}_g$ is then given by the line-of-sight projection of $\delta_g$ as\footnote{In practice, there is an additional magnification bias that arises due to the altered number of observed galaxies by gravitational lensing~\cite{Turner:1984ch, Villumsen:1995ar}. This effect can be incorporated by modifying the window function as
\begin{align}\label{magnification}
	W^{(i)}(\chi) \to W^{(i)}(\chi)+ (5s-2)\int_\chi^{\chi_*}\d\chi'\, W_\kappa(\chi')\frac{\d n_i}{\d \chi'}\, ,
\end{align}
which depends on the slope $s=\d\log N(<m_*)/\d m_*$ of the number count $N(m_*)$ as a function of the magnitude limit $m_*$~\cite{Hui:2007cu, Liu:2013yna}.}
\begin{align}
	\delta^{(i)}_g(\hat\n) &= \int_0^\infty\d\chi\, W^{(i)}_g(\chi)\delta_g(\chi,\chi\hat\n)\, ,\label{deltai}\\[3pt]
		W^{(i)}_g(\chi) &\equiv \frac{1}{\bar n_i}\frac{\d n_i}{\d \chi}\, , \quad \bar n_i =\int_0^\infty \d \chi\frac{\d n_i}{\d \chi}\, ,\label{Wi}
\end{align}
where $W^{(i)}_g$ is the normalized redshift distribution of galaxies in the $i$-th bin, $\d n_i/\d z = -H(z)\d n_i/\d\chi$.
We provide the specific form for the underlying redshift distribution for different experiments in Appendix~\ref{app:more}.

Observables of our interest are the expectation values of products of harmonic coefficients. For $n$ operators, this takes the form
 \begin{equation}
	\langle \O_{\ell_1 m_1}\cdots \O_{\ell_n m_n}\rangle = (4\pi)^ni^{\ell_{1\cdots n}}\!\left[\prod_{i=1}^n\int_0^\infty\!\d\chi_i\hs W_\O(\chi_i)\!\int_{\k_i}j_{\ell_i}(k_i\chi_i)Y_{\ell_im_i}^*(\hat\k_i)\right]\!\langle \widetilde\O_1\cdots \widetilde\O_n\rangle\, ,\label{angnpt}
\end{equation}
where $i^{\ell_{1\cdots n}}\equiv i^{\ell_1+\cdots+\ell_n}$, $\widetilde\O_i\equiv \widetilde\O_i(z_i,k_i)$, and $j_\ell$ is the spherical Bessel function. Inside the integrand is the Fourier-space $n$-point function
\begin{align}
	\langle \widetilde\O_1\cdots \widetilde\O_n\rangle = f_n(\{z_i,\k_i\}_{i=1}^n)\times (2\pi)^3\Ddelta(\k_1+\cdots+\k_n)\, ,
\end{align}
with the delta function being present as a consequence of momentum conservation. 
Spatial isotropy implies that $f_n$ is a function only of dot products of external momenta. 

The structure of the projection integral~\eqref{angnpt} crucially depends on the form of $f_n$. For correlations of matter overdensity, $f_n$ takes the form
\begin{align}
	f_n^{\rm G}(\{z_i,\k_i\}_{i=1}^n) &=  D_i^{p_i}\cdots  D_n^{p_n}\langle\delta_{p_1}(\k_1)\cdots\delta_{p_n}(\k_n)\rangle'\, ,\\[3pt]
	f_n^{\rm NG}(\{z_i,\k_i\}_{i=1}^n) &= \M_1\cdots \M_n\,\langle\phi(\k_1)\cdots\phi(\k_n)\rangle'\, ,
\end{align}
for Gaussian and non-Gaussian initial conditions respectively, with $M_i\equiv M(z_i,k_i)$ and $\delta_{p_i}$ the $p_i$-th order perturbative solution~\eqref{deltan}. To understand their momentum dependence, let us first consider the choices of $p_i$ that are required to form an $L$-loop diagram from a Gaussian initial condition. 
This can be done by counting the number of integrals and delta functions, which we denote by $I$ and $D$, respectively.  Each $p_i$-th order solution comes with $p_i$ density fields, $p_i$ integrals, and one delta function that imposes momentum conservation for the vertex.\footnote{For $p_i=1$ that corresponds to the linear solution, this implies one integral with a delta function, so the overall counting \eqref{L}, which involves taking the difference between $I$ and $D$, is unaffected.} We need an even number of fields to fully contract them, after which we get a product of power spectra and delta functions with half the number of fields contracted. This implies
\begin{align}
	I = \sum_i p_i\, , \quad  D =  \frac{1}{2}\sum_i p_i+n\, ,
\end{align}
assuming $\sum_i p_i=\text{even}$. Reserving one delta function for the total momentum conservation, we find that the condition for having an $L$-loop diagram is
\begin{align}
	L= 1-D+I= 1-n+\frac{1}{2}\sum_i p_i\, .\label{L}
\end{align}
It is straightforward to see that this gives us the familiar power spectra: $\{p_1,p_2\}=\{1,1\}$ for $L=0$ and $\{1,3\}$, $\{2,2\}$ for $L=1$.

Invariance under spatial rotations implies that the azimuthal dependence can always be factored out, after which the physical degrees of freedom are described by the multipoles corresponding to the edges and diagonals of an $n$-gon~\cite{Hu:2001fa, Mitsou:2019ocs, Lee:2020ebj}. This can be achieved by first performing all the angular integrations, which, however, can be a rather involved exercise for generic $n$, $L$.\footnote{As an example, consider the contribution $\{p_1,{\cdots}\hs,p_n\}=\{1,{\cdots}\hs,1,n-1\}$ to the $n$-point tree diagram
 \begin{align}
 	f_n^{\rm G}(\{z_i,\k_i\}_{i=1}^n) = (n-1)!D_1\cdots D_{n-1}D_n^{n-1} P_1\cdots P_{n-1} F_{n-1}(\k_1,\cdots,\k_{n-1})+\text{$(n-1)$ perms}\, ,
 \end{align}
where the SPT kernel $F_{n-1}$ is a rational function of dot products of its arguments.  
For $n\ge 4$, not all dot products can be expressed just in terms of external momentum magnitudes, which makes it nontrivial to integrate all the angular factors against the spherical harmonics. See \cite{Regan:2010cn, Smith:2015uia, Lee:2020ebj} for studies at $n=4$.} 
To demonstrate this procedure, let us consider the simplest case, for which (each term that contributes to) the Fourier-space correlator takes the following factorized form
\begin{align}
	\langle\widetilde\O_1\cdots \widetilde\O_n\rangle =g_1(z_1,k_1)\cdots g_n(z_n,k_n)\int_{\mathbb{R}^3}\d^3r\, e^{i\k_1\cdot\r}\cdots e^{i\k_1\cdot\r}\, ,\label{separable}
\end{align}
in terms of the external magnitudes $k_i$, where we have used the integral representation of the delta function.  

Using the plane-wave expansion and performing all the angular integrations, we arrive at
 \begin{align}
	\langle \O_{\ell_1 m_1}\cdots \O_{\ell_n m_n}\rangle &=\frac{\G^{\ell_1\cdots\ell_n}_{m_1\cdots m_n}}{(2\pi^2)^n}\int_0^\infty \d r\,r^2 I_{\ell_1}^{(1)}(r)\cdots I_{\ell_n}^{(n)} (r) \, , \label{angcorr}
 \end{align}
 where we have defined
 \begin{align}
\G^{\ell_1\cdots\ell_n}_{m_1\cdots m_n} &\equiv \int_{S^2}\d\Omega_{\hat\r}\,  Y_{\ell_1m_1}^*(\hat\r)\cdots Y_{\ell_n m_n}^*(\hat\r)\, , \label{geofator}\\[3pt]
I_\ell^{(i)}(r) &\equiv 4\pi\int_0^\infty\!\d\chi\hs W_\O(\chi)\!\int_0^\infty \d k\, k^2j_{\ell}(k\chi)j_{\ell}(kr)g_i(\chi,k)\, .\label{Iell}
 \end{align}
We see that for a separable ansatz in Fourier space, the angular integral conveniently factorizes.  The geometric factor \eqref{geofator} can be evaluated by the iterative use of the spherical harmonics product expansion
 \begin{align}
 	Y_{\ell_1m_1}Y_{\ell_2m_2}&=\sum_{\ell_3m_3}\G_{m_1m_2m_3}^{\ell_1\ell_2\ell_3}Y_{\ell_3m_3} = \sum_{\ell_3m_3} g_{\ell_1\ell_2\ell_3}\begin{pmatrix}
		\ell_1 & \ell_2 &\ell_3 \\ m_1 & m_2 & m_3
	\end{pmatrix}Y_{\ell_3m_3}\nn[5pt] 
 	&\equiv\sum_{\ell_3m_3}\sqrt{\frac{(2\ell_1+1)(2\ell_2+1)(2\ell_3+1)}{4\pi}}\begin{pmatrix}
		\ell_1 & \ell_2 &\ell_3 \\ 0&0&0
	\end{pmatrix}\begin{pmatrix}
		\ell_1 & \ell_2 &\ell_3 \\ m_1 & m_2 & m_3
	\end{pmatrix}Y_{\ell_3m_3}\, ,\label{YlmYlm}
 \end{align}
where round-bracket matrices denote the Wigner 3-$j$ symbols.  

For $n=2,3$, we have
\begin{align}
	\langle\O_{\ell m}\O_{\ell' m'}'\rangle &= \delta_{\ell\ell'}^{\rm K}\delta_{mm'}^{\rm K} C_\ell^{\O\O'}\, , ,\label{eq:powerspectrum}\\[3pt]
		\langle\O_{1,\ell_1 m_1}\O_{2,\ell_2 m_2}\O_{3,\ell_3 m_3}\rangle  &= \G^{\ell_1\ell_2\ell_3}_{m_1m_2m_3} b_{\ell_1\ell_2\ell_3}^{\O_1\O_2\O_3}\, ,\label{eq:bispectrum}
\end{align}
where $\delta^{\rm K}$ is the Kronecker delta. The physical degrees of freedom of the angular bispectrum are characterized by the reduced bispectrum $b_{\ell_1\ell_2\ell_3}^{\O_1\O_2\O_3}$, which is constrained by the triangle inequality $|\ell_i-\ell_j|\le \ell_k\le \ell_i+\ell_j$ for $i\ne j\ne k$ as well as the conditions $m_1+m_2+m_3=0$ and $\ell_1+\ell_2+\ell_3=\text{even}$.  

\subsection{Projection Integrals}\label{sec:gen}

In the previous section, we used separability in Fourier space to write the angular power spectra and bispectra in the general form \eqref{angcorr}. 
Evaluating the remaining momentum integral \eqref{Iell}, however, is numerically challenging due to the presence of spherical Bessel functions in the integral kernels.
Fortunately, certain approximations can be employed to greatly simplify the integrals. 
In this section, we review the Limber approximation and the FFTLog method for computing correlators in angular space, and explain how approximating integral kernels with polynomials can be useful for dealing with scale-dependent kernels.

\subsubsection{Limber Approximation}

A common approach to evaluate the projection integrals is to use the Limber approximation~\cite{Limber:1954zz}. This amounts to replacing the spherical Bessel functions as delta functions, $j_\ell(x)\simeq \sqrt{\frac{\pi}{2\ell}}\Ddelta(\ell+\frac{1}{2}-x)$, in which case the two-Bessel integral can be approximated as~\cite{Limber:1954zz, LoVerde:2008re, DiDio:2018unb}
\begin{align}\label{limber}
	\int_0^\infty \d k\, k^2j_\ell(k\chi)j_{\ell'}(k\chi')f(k)\, &\simeq\, \frac{\pi}{2\chi^2}f(\ell/\chi) \times\begin{cases} \displaystyle \Ddelta(\chi-\chi'); & \ell'=\ell\, , \\[10pt] \displaystyle \frac{\ell+1/2}{\ell}\,\Ddelta(\tfrac{\ell+1}{\ell}\chi-\chi'); & \ell'=\ell+1 \, , \end{cases}
\end{align}
in the limit $\ell\gg 1$, where we have dropped subleading terms in $1/\ell$. 
With this approximation, the angular power spectrum is dramatically simplified to
\begin{align}
	C_\ell^{\O\O'} = \int_0^\infty \frac{\d\chi}{\chi^2}\, W_\O(\chi)W_{\O'}(\chi)P^{\O\O'}\!(\chi,\chi,\ell/\chi)\, .
\end{align}
The validity of the Limber approximation for angular spectra at small angular scales has been well studied, see e.g.~\cite{Lemos:2017arq,Deshpande:2020jjs,Bellomo:2020pnw,Bernal:2020pwq}.\footnote{An interesting counter-example is provided by the non-Gaussian covariance of the power spectrum (corresponding to the collapsed limit of the angular trispectrum), for which the Limber approximation fails even for high multipoles~\cite{Lee:2020ebj}. This is due to the intrinsic error of the approximation being larger than the precision required for accurate cancellation amongst different contributions.} 
However, this approximation also has a number of known drawbacks: it fails for low multipoles, narrow tomographic bins, and tomographic cross-spectra. Since these might become relevant for future surveys, it is desired to have a more accurate numerical method for computing angular spectra.

\subsubsection{FFTLog}

The prime difficulty in computing angular spectra has to do with the presence of highly-oscillatory Bessel functions in the projection integral~\eqref{Iell}, which makes its evaluation rather complicated with a brute-force numerical method. Instead, the FFTLog algorithm~\cite{Assassi:2017lea, Gebhardt:2017chz, Schoneberg:2018fis} provides an efficient means to evaluate the angular spectra.

The FFTLog of a function is defined essentially as its discrete Fourier transform in log-$k$ space. This decomposes the function into a sum of complex power laws instead of the usual plane waves. 
Taking the FFTLog of $f(\chi,k)$ over a finite interval $[k_{\rm min},k_{\rm max}]$, we get
\begin{align}\label{eq:FFTLog}
	f(\chi,k) \simeq \sum_{n=-N_\eta/2}^{N_\eta/2}c_n(\chi) k^{-b+i\eta_n}\quad\text{with} \quad \eta_n\equiv \frac{2\pi n}{\log(k_{\rm max}/k_{\rm min})}\, ,
\end{align}
where the coefficients $c_m$ are given by the inverse transform
\begin{align}\label{FFTco}
	c_n(\chi)=\frac{2-\delta^{\rm K}_{|n|,N_\eta/2}}{2N_\eta}\sum_{m=0}^{N_\eta-1}f(\chi,k_m)k_m^{b}k_{\rm min}^{-i\eta_n}e^{-2\pi i mn/N_\eta}\, .
\end{align} 
Some care needs to be taken when using this method: since the transform is over a finite interval with a finite number of sampling points, not all functions have the same convergence properties. Notice above that we have implicitly taken the FFTLog of $k^{b}f(\chi,k)$ instead of $f(\chi,k)$, with some $b\in\mathbb{R}$. This extra parameter $b$ is in principle arbitrary but can be appropriately fixed to obtain a more convergent result.

The virtue of FFTLog is that the contribution of each summand to the projection integral can now be computed analytically. For example, taking the FFTLog of $g_i(\chi,k)$ in \eqref{Iell} gives
\begin{align}\label{eq:Iellr}
I_\ell^{(i)}(r) \simeq \sum_n  \int_0^\infty \d\chi\,  c_n(\chi) \W_\delta(\chi)\chi^{-\nu_n} {\sf I}_\ell (\nu_n,\tfrac{r}{\chi})\, ,
\end{align}
where we have defined $\nu_n\equiv 3-b+i\eta_n$ and
\begin{align}
	{\sf I}_\ell(\nu,w) &\equiv 4\pi\int_0^\infty \d x\, x^{\nu-1}j_\ell(x)j_\ell(wx)\nn
	&= \frac{2^{\nu-1}\pi^2\Gamma(\ell+\frac{\nu}{2})}{\Gamma(\frac{3-\nu}{2})\Gamma(\ell+\frac{3}{2})}\,w^\ell\, {}_2F_1\Bigg[\begin{array}{c} \frac{\nu-1}{2},\hs \ell+\frac{\nu}{2} \\[2pt] \ell+\frac{3}{2}\end{array}\Bigg|\, w^2\Bigg]\quad (|w| \le 1)\, ,
	\label{WSint}
\end{align}
with ${}_2F_1$ the Gauss hypergeometric function. The answer for $|w| > 1$ can be obtained by ${\sf I}_\ell(\nu,w) = w^{-\nu} {\sf I}_\ell(\nu,\tfrac{1}{w})$. 
Note that the formula \eqref{WSint} is {\it exact}, meaning that unlike the Limber approximation the calculation is valid for any $\ell$, as long as the expansion in \eqref{eq:FFTLog} converges in the first pace. Another nice feature of FFTLog is that ${}_2F_1$ has well-known analytic properties and admits fast-converging series representations, suitable for numerical evaluation.
Efficient algorithms for evaluating the Gauss hypergeometric function in this context are provided in detail in~\cite{Assassi:2017lea, Gebhardt:2017chz}.

\subsubsection{Polynomial Approximation} \label{sec:poly}

In order for the above method to work, it is necessary that the $\chi$- and $k$-dependent parts of the projection integral are factorized, so that the momentum integral can be done analytically.
While this can be accomplished by allowing the coefficients $c_m$ to be $\chi$ dependent as in \eqref{eq:FFTLog}, we will find it more convenient to impose factorization first and then take the FFTLog of the purely $k$-dependent part. 

\begin{figure}[t]
    \centering
                  \includegraphics[width=\textwidth]{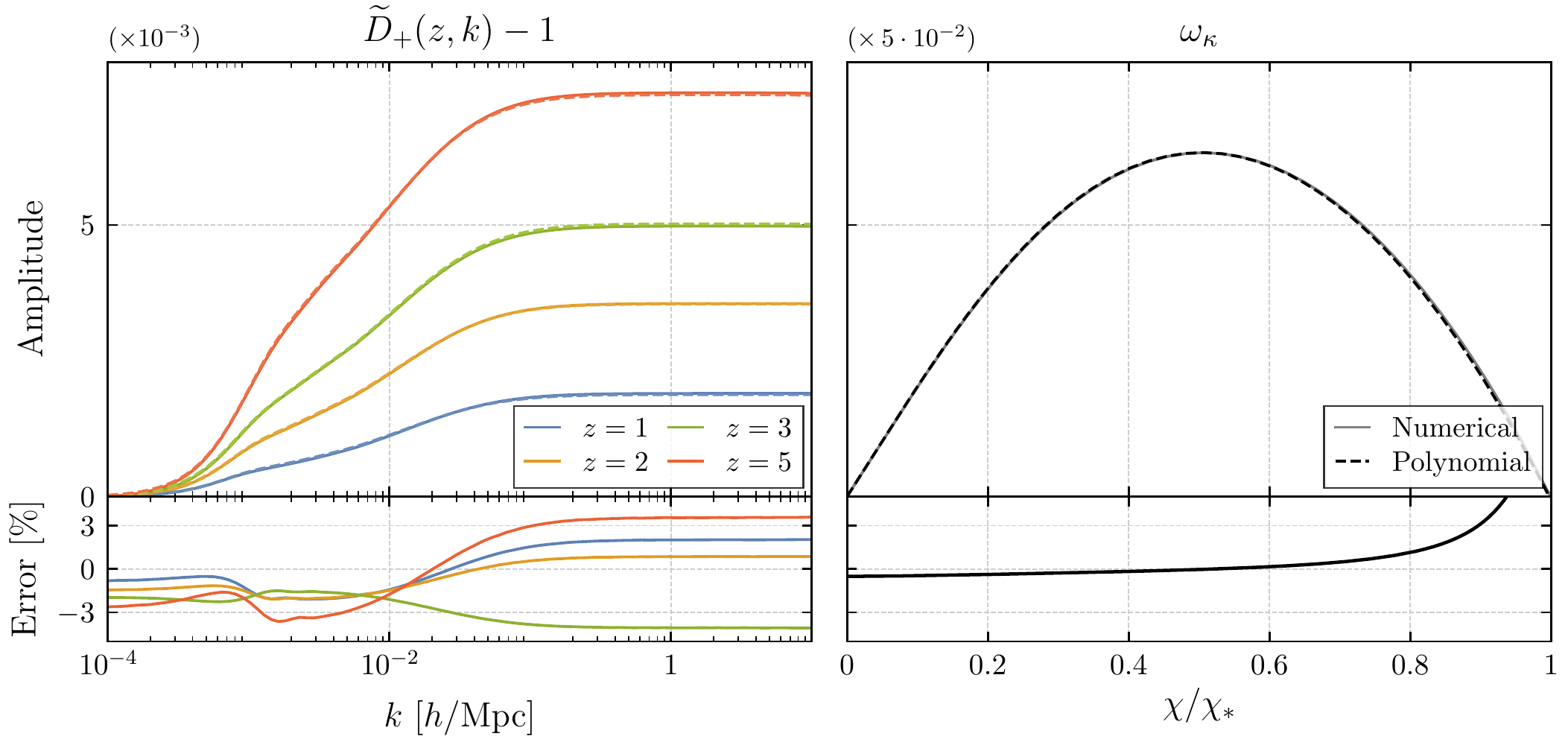}  
         \\
    \caption{Polynomial approximation of the growth function and the CMB lensing kernel, with $N_{\rm poly}=3$ and $14$, respectively. In the left panel, we show the normalized growth function defined by $\widetilde D_+(z,k)\equiv D_+(z,k)/D_+(z,0)$ for different redshifts. In the right panel, we show the normalized CMB lensing kernel $\omega_{\kappa}\equiv (1+z)\chi(\chi_*-\chi) D_+(z,0)/\chi_*^{2}$. The relative errors due to the polynomial approximation are shown in the lower panels.}
    \label{fig:poly}
\end{figure}

For the power spectrum with massive neutrinos, the $k$-$\chi$ coupling comes from the linear growth function $D_+$ as well as its logarithmic derivative $f_+$ present in the RSD term. Since these are relatively smoothly varying functions of $\chi$ for a fixed $k$, it is possible to approximate them with a low-degree polynomial in $\chi$. 
To this end, we employ the following polynomial approximation scheme:\footnote{See \cite{Levi:2016tlf} for a similar prescription of using a polynomial approximation in the presence of massive neutrinos, taken at the level of the density field.}
\begin{align}
	F(k,\chi) &\simeq \sum_{p=0}^{N_{\rm poly}} w_p^F(k)\chi^p\, ,\label{poly}
\end{align}
where the function $F$ will be some combination of $D_+$, $f_+$, and the window function $W_\O$; for $F=W_\kappa$, its coefficients will be $k$ independent. 
To demonstrate the validity of this approximation, we show in Fig.~\ref{fig:poly} polynomial fits to $D_+$ as a function of $k$ for different redshifts and $W_\kappa$ as a function of $\chi$. We see that even cubic polynomials are sufficient to keep the error level down to $O(1\%)$ for $D_+$, while $O(10)$ terms is needed for a percent-level convergence of $W_\kappa$.

The expansion \eqref{poly} allows us take the FFTLog of a purely $k$-dependent function, which usually involves the combination $w_i^F(k)P^{\O\O'}\!(k)$. 
As we explain in Appendix~\ref{app:analytic}, the remaining calculation then essentially boils down to evaluating integrals of the form
\begin{equation}
	\int\d \chi\, \chi^{\alpha-1}\int\d \chi' \chi'^{\beta-1} \I_\ell(\nu,\tfrac{\chi}{\chi'})\, ,\label{eq:poly}
\end{equation}
with $\alpha,\beta\in\mathbb{C}$. Since $\I_\ell$ is given by the hypergeometric ${}_2F_1$, integrating it against power-law functions is straightforward, and gives the result in terms of the generalized hypergeometric function ${}_3F_2$---another numerically-friendly special function. This provides a fully analytic way of computing angular power spectra, which in general outperforms numerical integration methods for doing the $\chi$-$\chi'$ integrals. We refer the interested reader to Appendix~\ref{app:analytic} for technical details of this approach.

\subsection{Angular Polyspectra}\label{sec:angcorr}
In this section, we further elaborate on the FFTLog method and describe its application to our main observables: angular power spectra and bispectra.

\subsubsection{Angular Power Spectra}

\begin{figure}[t!]
    \centering
         \includegraphics[width=\textwidth]{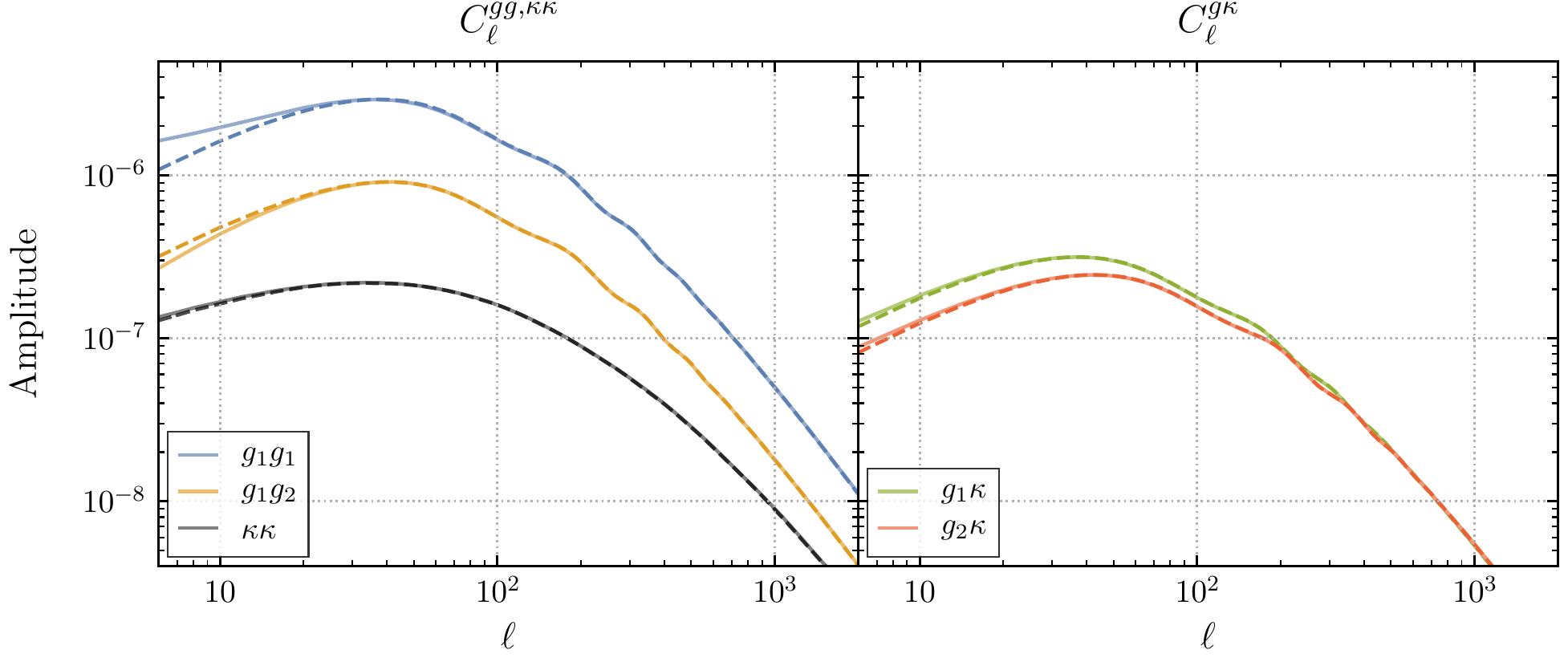}      
         \\
    \caption{Shapes of tree power spectra computed using FFTLog (solid lines) and the Limber approximation (dashed lines). We set $b_{\delta}=1$ and use the Gaussian window function with $\bar z=1,\, 1.2$ and $\sigma_{z}=0.1$ for galaxy overdensities denoted by $g_1$ and $g_2$, respectively.}
    \label{fig:ClComp}
\end{figure}

The angular power spectrum, defined in \eqref{eq:powerspectrum}, can be expressed as
\begin{align}
	C_\ell^{\O\O'} &= \frac{1}{2\pi^2}\int_0^\infty\d\chi\, W_\O(\chi)\int_0^\infty\d\chi'\, W_{\O'}(\chi') I_\ell^{\O\O'}\!(\chi,\chi';0)\, ,\label{Cell}\\
	I_{\ell\ell'}^{\O\O'}(\chi,\chi';n) &\equiv 4\pi\int_0^\infty\d k\, k^{2+n} j_\ell(k\chi)j_{\ell'}(k\chi')P^{\O\O'}\!(\chi,\chi',k)\, , 
\end{align}
with $I_{\ell}^{\O\O'}\equiv I_{\ell\ell}^{\O\O'}$. As described in the previous section, the computation can be proceeded with either FFTLog or Limber's method. As an illustration, Fig.~\ref{fig:ClComp} shows a comparison of the angular power spectra of galaxies and CMB lensing using the two methods, using a Gaussian window function $W_g \propto e^{-(z-\bar z)^{2}/2\sigma_z^2}$ at $\bar z=1,1.2$.
Although the two methods appear to agree very well for the large chosen width $\sigma_z=0.1$, it is apparent that the Limber-approximated galaxy power spectra start to deviate from the exact ones computed with FFTLog at low $\ell$. This choice of $\sigma_z$ reflects the expected size of redshift measurement errors for upcoming photometric surveys such as LSST (See Appendix~\ref{app:more}). For CMB lensing, the Limber approximation remains highly accurate due to its very wide window function. We will quantify the difference between the two methods in parameter estimation in Section~\ref{sec:fisher}.

It is straightforward to implement the linear Kaiser effect~\eqref{RSD} in angular space by noting that $\mu\to ik^{-1}\partial_\chi$.\footnote{In the plane-parallel (distant-observer) approximation, the mapping between the redshift- and real-space coordinates, $\s$ and $\x$, is given by $\s =\x -f_+ (\u\cdot\hat\n)\hat\n$, where $\u\equiv -\v/\H f$. Using this relation and taking the Fourier transform of the redshift-space density field gives~\cite{Scoccimarro:1999ed}
\begin{align}
	\delta_s(\k) = \int\d^3x\, e^{i\k\cdot\x}e^{-if_+ u\k\cdot\hat\n}(\delta(\x)+f\hat\n\cdot\nabla(\u\cdot\hat\n))\, .
\end{align}
This is a fully nonlinear density field in the plane-parallel approximation, whose leading-order expansion gives the Kaiser effect. The higher-order terms can be treated perturbatively, and its implementation in angular space was studied recently in~\cite{Gebhardt:2020imr}.} 
For example, the coupling integral for the galaxy-matter power spectrum at tree level in the presence of the RSD is
\begin{align}
	I_\ell^{gm}(\chi,\chi';0) = 4\pi\int_0^\infty\d k\,k^{2}\big(b_\delta\hs j_\ell(k\chi)-f_+(k,\chi) j_\ell''(k\chi)\big)j_\ell(k\chi')P^{mm}(\chi,\chi',k)\, .
\end{align}
We may express $j_\ell''$ as a linear combination of $j_\ell$ and $j_{\ell+1}$, after which we get
\begin{align}
	I_\ell^{gm}(\chi,\chi';0) &=(1+b_\delta)I_\ell^{mm}(\chi,\chi';0)-\frac{\ell(\ell-1)}{k^2\chi^2}I_\ell^{mm}(\chi,\chi';-2)+\frac{2}{k\chi}I_{\ell+1,\ell}^{mm}(\chi,\chi';0)\, .
\end{align}
Making a similar replacement for $P^{gg}$ results in a total of 9 terms. Another way of dealing with the RSD, which happens to be more convenient for window functions with vanishing contributions at the boundaries, is to integrate by parts so that the derivatives of the Bessel functions instead act on the $\chi$ integrand. Doing this requires first putting the $\chi$ and $k$ integrals in a factorized form, which can be achieved with the polynomial approximation described in the previous section. 

\subsubsection{Angular Bispectra}

Many physical tree bispectra can be expressed as a sum over separable terms of the form~\eqref{separable}. The reduced angular bispectrum can then be represented as
\begin{align}
	b_{\ell_1\ell_2\ell_3}^{\O_1\O_2\O_3} &= \sum_{p_1p_2p_3p_4}\sum_{n_1n_2n_3}c_{n_1n_2n_3}^{\O_1\O_2\O_3} \nn
	\times&\int_0^\infty \d r^2\, r^2 I_{\ell_1}^{\O_1\O_3}(r;n_1,p_1,p_3)I_{\ell_2}^{\O_2\O_3}(r;n_2,p_2,p_4) I_{\ell_3}^{\O_3}(r;n_3,p_3,p_4)+ \text{2 perms} \, ,\label{b123}
\end{align}
where $c_{n_1n_2n_3}^{\O_1\O_2\O_3}$ are constants and
\begin{align}
	I_{\ell}^{\O\O'}(r;n,p,q) & \equiv 4\pi\int_0^\infty\d\chi\, \chi^{p}\,\W_\O(\chi)\int_0^\infty\d k\, k^{2+n}j_{\ell}(kr)j_{\ell}(k\chi) \omega_{p}(k) \omega_{q}(k)  P^{\O\O'}\!(k)\, ,\label{Iell2}\\
	I_{\ell}^\O(r;n,p,q) & \equiv 4\pi\int_0^\infty\d\chi\, \chi^{p+q}\,\W_\O(\chi)\int_0^\infty\d k\, k^{2+n} j_{\ell}(kr)j_{\ell}(k\chi) \, .\label{Iell3}
\end{align}
Note that we have expanded the scale-dependent growth function as $D(k,\chi) = \sum_{p}\omega_{p}(k)\chi^{p}$ to preserve the desired separable form of the bispectra, as explained in Section~\ref{sec:poly}.
We have also defined the dressed window function
\begin{align}\label{WO}
	\W_\O(\chi) \equiv \begin{cases}
		b_\Q(\chi) W_g(\chi) & \O=\delta_g\, ,\ \Q\in\{\delta,\delta^2,\G_2\} \, ,\\ W_\kappa(\chi) & \O=\kappa\, ,
	\end{cases}
\end{align}
that includes redshift-dependent bias parameters. For Gaussian initial conditions, this can be done by expressing the Fourier-space bispectra \eqref{Bmmm}--\eqref{Bggg} in a separable form using
\begin{align}
	L_2(\k_1,\k_2) &= -\frac{1}{2}+\frac{k_1^2}{4k_2^2}+\frac{k_2^2}{4k_1^2}-\frac{k_3^2}{2k_1^2}-\frac{k_3^2}{2k_2^2}+\frac{k_3^4}{4k_1^2k_2^2}\, , \\[3pt]
	F_2(\k_1,\k_2)&=\frac{5}{14}-\frac{5}{28}\left(\frac{k_1^2}{k_2^2}+\frac{k_2^2}{k_1^2}\right)+\frac{3}{28}\left(\frac{k_3^2}{k_1^2}+\frac{k_3^2}{k_2^2}\right)+\frac{k_3^4}{14k_1^2k_2^2}\, .
\end{align}
Due to high/low powers of each momentum, the individual $k$-integrals may naively seem divergent in the ultraviolet/infrared regimes, even though a careful regularization will render the final bispectrum finite. These spurious divergences arise due to the fact that we have changed the order of integrations so that the radial integration that imposes momentum conservation is performed last,\footnote{Recall that the outer $r$ integral is precisely the radial part of the integral representation of the momentum-conserving delta function; c.f.~\eqref{separable}.} as a result of which we are keeping track of unphysical momentum configurations in the intermediate steps~\cite{DiDio:2016gpd}.

\begin{figure}[t!]
    \centering
         \includegraphics[width=\textwidth]{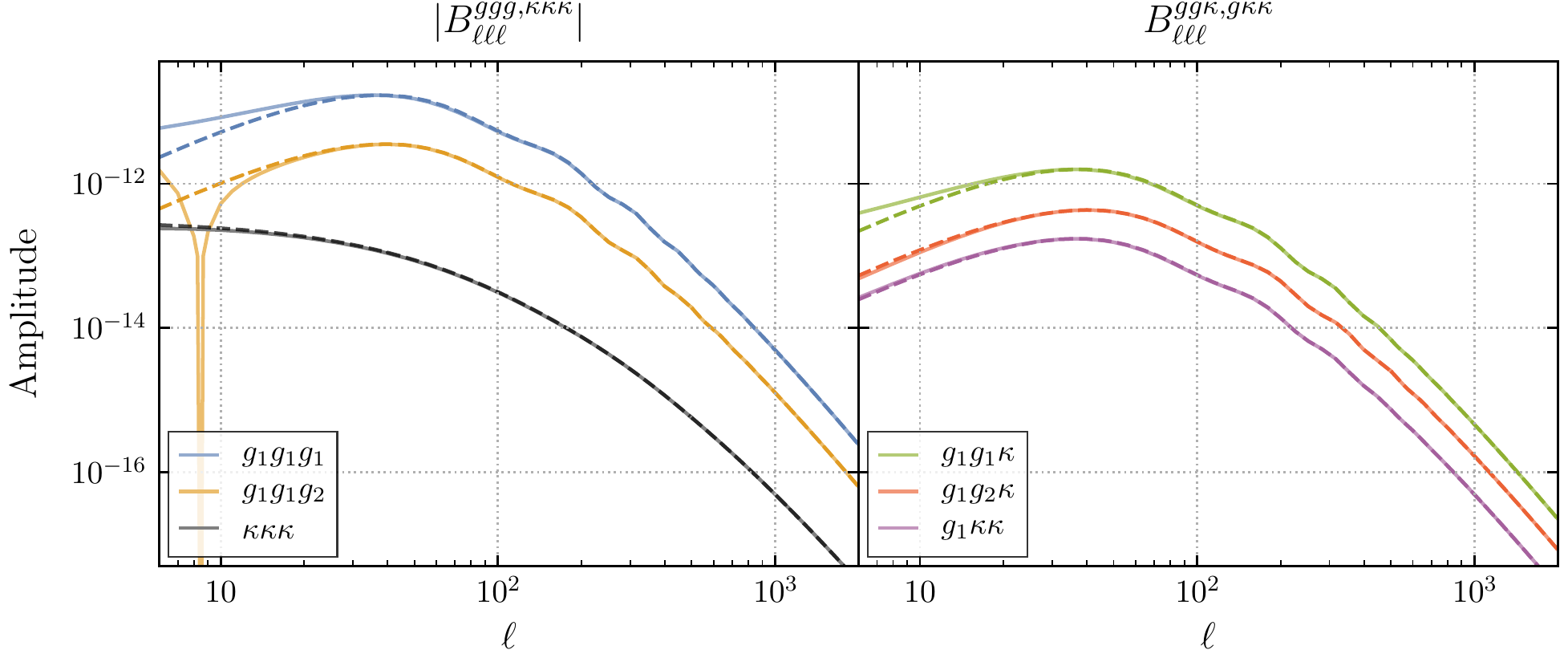}          
         \\
    \caption{Shapes of tree bispectra in equilateral configurations computed using  FFTLog (solid lines) and the Limber approximation (dashed lines). We set $b_{\delta}=1$, $b_{\delta^{2}}=b_{\G_{2}}=0$, and use the Gaussian window function with $\bar{z}=1,\, 1.2$ and $\sigma_{z}=0.1$ for galaxy overdensities denoted by $g_1$ and $g_2$, respectively.} 
    \label{fig:BlllComp}
\end{figure}

To deal with this issue, it is convenient to shift powers of momenta between different integrals prior to numerical evaluation by the use of the differential operator~\cite{Assassi:2017lea, Schoneberg:2018fis}
\begin{align}
	 \D_\ell(r)\equiv -\partial_r^2-\frac{2}{r}\partial_r+\frac{\ell(\ell+1)}{r^2}\, ,\quad \D_\ell(r)j_\ell(kr) =k^2j_\ell(kr)\, .\label{diffop}
\end{align}
For example, we can lower two powers of $k_1$ by integrating this operator by part as
\begin{align}
	&\int_0^\infty \d r^2\, r^2 I_{\ell_1}^{\O_1\O_3}(r;n_1)I_{\ell_2}^{\O_2\O_3}(r;n_2) I_{\ell_3}^{\O_3}(r;n_3) \nn
	&\qquad = \int_0^\infty \d r^2\, r^2 I_{\ell_1}^{\O_1\O_3}(r;n_1-2) \D_{\ell_1}(r)\Big[I_{\ell_2}^{\O_2\O_3}(r;n_2) I_{\ell_3}^{\O_3}(r;n_3)\Big]\, .\label{ibp}
\end{align}
The radial derivative operator $\D_\ell(r)$ can then be taken analytically by
\begin{align}
	\D_{\ell}(r)\Big[I_{\ell_2}^{\O_2\O_3}(r;n_2) I_{\ell_3}^{\O_3}(r;n_3)\Big] &= I_{\ell_2}^{\O_2\O_3}(r;n_2+2) I_{\ell_3}^{\O_3}(r;n_3)+I_{\ell_2}^{\O_2\O_3}(r;n_2) I_{\ell_3}^{\O_3}(r;n_3+2)\nn[3pt]
	&\hskip -50pt -\frac{\ell(\ell+1)}{r^2}I_{\ell_2}^{\O_2\O_3}(r;n_2) I_{\ell_3}^{\O_3}(r;n_3)-2\partial_r I_{\ell_2}^{\O_2\O_3}(r;n_2) \partial_r I_{\ell_3}^{\O_3}(r;n_3)\, ,
\end{align}
and using the identity $\partial_r j(kr)=\frac{\ell}{r} j_\ell(kr)-kj_{\ell+1}(kr)$ for the linear derivative term (see Appendix A of~\cite{Lee:2020ebj}).\footnote{In~\cite{Assassi:2017lea, Schoneberg:2018fis}, they instead considered $\D_\ell(\chi)$ on $j_\ell(k\chi)$ and integrated it by part to act on the window function as
\begin{align}
	I_\ell^{\O\O'}(r;n) = 4\pi\int_0^\infty\d\chi\, [\widetilde\D_\ell W_\O](\chi)\int_0^\infty\d k\, k^{n} j_\ell(k\chi)j_\ell(kr) D_+(\chi,k) P^{\O\O'}\!(k) +\text{BT}\, ,
\end{align}
where `BT' denotes boundary terms and
\begin{align}
	\widetilde\D_\ell(\chi) \equiv -\partial_\chi^2 + \frac{2}{\chi}\partial_\chi + \frac{\ell(\ell+1)-2}{\chi^2} = \D_\ell(\chi) +\frac{4}{\chi}\partial_\chi -\frac{2}{\chi^2}\, .
\end{align}
This representation is useful for smooth window functions with vanishing boundary terms. Note, in contrast, that there are no boundary terms in \eqref{ibp}.} 
Because the $r$ integrand is a smoothly-varying function for typical window functions, the derivatives may also be taken numerically.

A comparison of the bispectra of galaxies and CMB lensing using the Limber approximation and with the FFTLog method is shown in Fig.~\ref{fig:BlllComp}, for equilateral configurations ($\ell_1=\ell_2=\ell_3$) and $b_\delta=1$, $b_{\delta^{2}}=b_{\G_{2}}=0$. As with the power spectrum case, the Limber approximation works well for CMB lensing, but it breaks down for galaxy spectra at low $\ell$. In particular, the galaxy cross-spectrum flips its sign at low $\ell$, which is not captured by the Limber approximation. More details on how these differences affect parameter constraints will be given in Section~\ref{sec:fisher}.

Similar considerations hold for angular bispectra from non-Gaussian initial conditions. The matter bispectrum for the local-type non-Gaussianity---Eq.~\eqref{Bphi} with $K_{\rm NL}=1$---is in a manifestly separable form. The corresponding angular bispectrum is then given by
\begin{align}
	b_{\ell_1\ell_2\ell_3}^{\O_1\O_2\O_3} &= 2\fnl \int_0^\infty \d r^2\, r^2 \tilde I_{\ell_1}^{\O_1\O_1}(r;0)\tilde  I_{\ell_2}^{\O_2\O_2}(r;0) \tilde I_{\ell_3}^{\O_3}(r;0)+ \text{2 perms} \, ,\label{b123PNG}
\end{align}
where
\begin{align}
	\tilde I_\ell^{\O\O}(r;n) & = 4\pi\int_0^\infty\d\chi\, \widetilde\W_\O(\chi)\int_0^\infty\d k\, k^{2+n} j_\ell(kr)j_\ell(k\chi) M(\chi,k) P^{\phi\phi}\!(k)\, ,\label{Iell4}\\
	\tilde I_\ell^\O(r;n) & = 4\pi\int_0^\infty\d\chi\, \widetilde\W_\O(\chi)\int_0^\infty\d k\, k^{2+n} j_\ell(kr)j_\ell(k\chi) M(\chi,k)\, ,\label{Iell5}
\end{align}
and $\widetilde\W_\O$ is the same as $\W_\O$ in \eqref{WO} with the restriction $\Q=\delta$. We see that \eqref{b123PNG} has the same structure as \eqref{b123} modulo the growth function dependence.

\vskip 10pt
\begin{tcolorbox}[title=Summary of new results]
\begin{itemize}
		\item Applied the polynomial approximation \eqref{poly} to scale-dependent line-of-sight kernels. For each tomographic bin, as less as three terms are typically sufficient to reach convergence at the level of $O(1\%)$, while the CMB lensing kernel requires about 15 terms.
		\item Derived fully analytic formulas for computing angular power spectra by combining the polynomial approximation and FFTLog. These are expressed in terms of the generalized hypergeometric functions ${}_3F_2$ and the details are provided in Appendix~\ref{app:analytic}.
\end{itemize}
\end{tcolorbox}

\section{Parameter Forecasts}\label{sec:fisher}

In this section, we use the standard Fisher methodology to forecast constraints on cosmological parameters. We first provide the details of the Fisher analysis of angular power spectra and bispectra in~\S\ref{sec:model}. 
We then present our forecast results for LSST and CMB-S4 in \S\ref{sec:res}. The experimental specifications and the results for other galaxy surveys can be found in Appendix~\ref{app:more}.

\subsection{Fisher Methodology}\label{sec:model}

\begin{table}[b!]
\centering
\begin{tabular}{ c l c c}
\toprule
Parameter & Meaning & Fiducial\\
\midrule 
$H_0$ & Hubble parameter at $z=0$ [km/s/Mpc] &  67.4\\ 
$\omega_c$ & Cold dark matter density at $z=0$, $\omega_c=\Omega_{c0} h^2$ &  0.120\\ 
$\omega_b$ & Baryon density at $z=0$, $\omega_b=\Omega_{b0} h^2$ & 0.0224\\ 
$10^9A_s$ & Scalar amplitude at $k_\star=0.05\hs\text{Mpc}^{-1}$ &  2.100\\ 
$n_s$ & Spectral tilt & 0.965\\ 
$\tau$ & Optical depth & 0.054\\ 
\midrule
$M_\nu$ & Total neutrino mass [meV] & 60\\ 
$\fnl$ & Local non-Gaussianity amplitude &  0\\ 
\bottomrule
\end{tabular} 
\caption{List of parameters of the reference 
$\nu\Lambda\text{CDM}{+}\fnl$ cosmology and their fiducial values, taken from~\cite{Aghanim:2018eyx}.}
\label{tab:param}
\end{table} 

We use the Fisher matrix formalism~\cite{Tegmark:1997rp} to obtain constraints on cosmological parameters and analyze the degeneracies between them. 
We consider the following set of parameters in our analysis:
\begin{equation}\label{eq:lambda}
	\boldsymbol{\lambda} = \underbrace{\{H_0,\omega_c,\omega_b,A_s,n_s,\tau\} }_{\Lambda\text{CDM}}\, \cup\,  \underbrace{\{M_\nu, \fnl \} }_{\text{non-minimal}}\,\cup\, \bigcup_{i=1}^{N_z} \underbrace{\{\bar b_\delta^{(i)},\bar b_{\delta^2}^{(i)},\bar b_{\G_2}^{(i)},\bar b_{\partial^2\delta}^{(i)}\} }_{\rm nuisance}\, .
\end{equation}
The physical cosmological parameters---the six parameters of the flat $\Lambda$CDM cosmology and two additional parameters for non-minimal scenarios---and their fiducial values are summarized in Tab.~\ref{tab:param}. 
The index $i$ for the nuisance parameters runs over tomographic bins, $1,\cdots\bhs ,N_z$, and $\bar b_\O^{(i)}$ parameterizes the amplitude of the bias operator $\O$ in the $i$-th tomographic bin. More precisely, the tomographic bias coefficients are expressed as $ \bar b_\O^{(i)} b_\O^{(i)}(z)$, where $b_\O^{(i)}(z)$ are fixed according to some fiducial redshift dependence, while $\bar b_\O^{(i)}$ are treated as free parameters.
For $b_\delta$, we fix its fiducial value according to the model chosen for each experiment. For the higher-order biases, we use the fitting formulas obtained from $N$-body simulations~\cite{Lazeyras:2015lgp} to write them as functions of the linear bias; see Appendix~\ref{app:bias} for details. 
Note that the bias $b_{\Gamma_{3}}$ is not present in~\eqref{eq:lambda}: it turns out that this is nearly degenerate with the other parameters (see e.g.~\cite{Ivanov:2019pdj, DAmico:2019fhj}), and hence we choose not to vary it. 
Although the bias/EFT parameters are interesting from the perspective of understanding the nonlinear galaxy formation process, we will eventually marginalize over them as our main interest is in obtaining constraints on the cosmological parameters. 

In our Fisher analysis, we will only consider the leading contributions to the covariance matrix by only keeping its Gaussian (disconnected) part; that is, we assume that the density fields are Gaussian in the noise. This means that we neglect higher-order terms from connected four- and six-point correlation functions that give subleading contributions to the covariance in \eqref{F2pt} and~\eqref{F3pt}, respectively. These corrections have not yet been fully computed for galaxy statistics in the harmonic space (but see~\cite{Lacasa:2017ufk, Lee:2020ebj}), and their impact on parameter constraints remains to be studied.

\subsubsection{Power Spectrum}
Here we present the details of the Fisher matrix for angular power spectra. 
LSST has an effective redshift coverage of $[0,7]$, which we split into $N_z$ bins. The set of observables is then
\begin{align}
	\O\in \{\delta_g^{(1)},\delta_g^{(2)},\cdots,\delta_g^{(N_z)},\kappa\}\, .
\end{align}
For the power spectrum, we then have $N_z+1$ auto-spectra, $M$ $gg$ cross-spectra (with $M$ depending on the overlap between different window functions, to be specified), and $N_z$ $g\kappa$ cross-spectra; this gives a total of $2N_z+M+1$ two-point observables to consider.

\begin{figure}[t!]
    \centering
         \includegraphics[width=\textwidth]{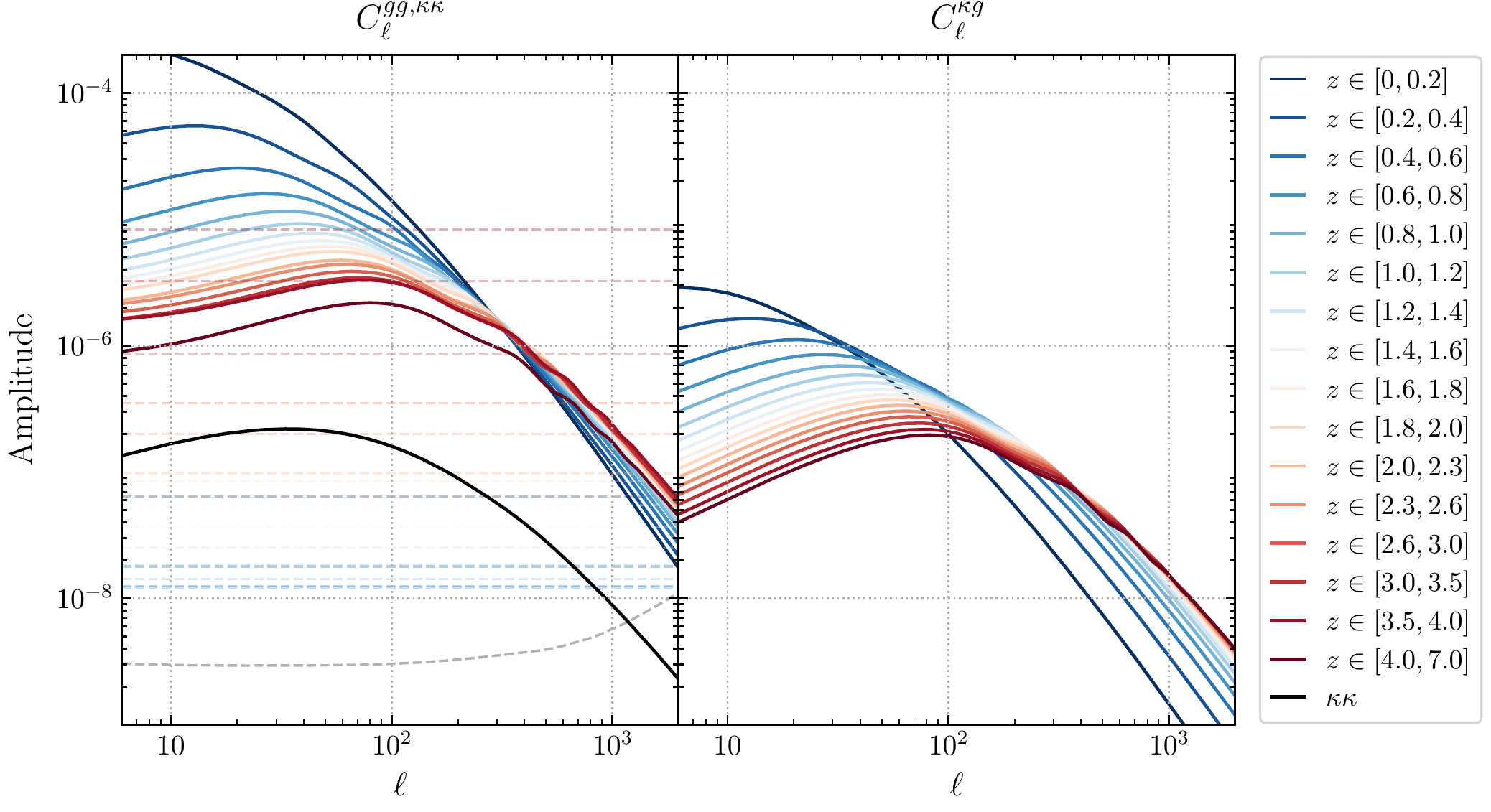}
    \caption{Angular power spectra of the reference cosmology with LSST window function of 16 tomographic bins. The solid lines show the galaxy and CMB lensing auto- and cross-spectra at tree level. The horizontal dashed lines show the shot noise corresponding to different tomographic bins. We use the FFTLog method to compute the power spectra up to $\ell_{\rm L}$ that corresponds to $k_{\rm L}=0.05\hMpc$, and use the Limber approximation for $\ell>\ell_{\rm L}$. The dashed gray line shows the lensing reconstruction noise for CMB-S4.}
      \label{fig:ClLSST}
\end{figure}

In our main forecast, we use the same redshift bin decomposition as in~\cite{Yu:2018tem} with edges given by the set $\{0, 0.2, 0.4, 0.6, 0.8, 1, 1.2, 1.4, 1.6, 1.8, 2, 2.3, 2.6, 3, 3.5, 4, 7\}$. 
We then take photometric redshift errors of LSST into account by convolving each non-overlapping tomographic bin with a Gaussian error kernel with width $\sigma_z=0.05(1+z)$; see Fig.~\ref{fig:ClLSST} and Appendix~\ref{app:more} for details. These redshift errors induce nonzero overlaps between different tomographic bins, resulting in non-negligible tomographic cross-spectra.
For $gg$ cross-spectra, we find it sufficient to include correlations between $\delta_g^{(i)}$ and $\delta_g^{(j)}$ with $|i-j|\leq2$, i.e.~those between adjacent bins and next-to-adjacent ones. In this case, there are $M=2N_z-3$ such cross-spectra, and the total number of two-point observables (including $\kappa\kappa,g\kappa$ and $gg$ auto-spectra) that we consider is $4N_z-2 = 62$.

We model the covariance between two power spectra as
\begin{align}
	{\sf C}^{\X\Y,\X'\Y'}_\ell 	&=\frac{1}{f_{\rm sky}}\frac{1}{2\ell+1}\Big[	(C^{\X\X'}_\ell+\delta_{\X\X'}N^{\X}_\ell)(C^{\Y\Y'}_\ell+\delta_{\Y\Y'}N^{\Y}_\ell)+(\X'\leftrightarrow \Y')\Big]\,,
\end{align}
where $\X,\Y\in\O$ and $f_{\rm sky}$ is the sky fraction observed. The noise spectra $N_\ell^\X$ contribute only for auto-spectra, and it is given by either the lensing reconstruction noise for $\X=\kappa$ or the shot noise
\begin{align}
	\N_\ell^{\delta_g^{(i)}} =  \left(\frac{\d n_i}{\d z}\right)^{-1} ,
\end{align}
where $\d n_i/\d z$ is the galaxy number density in the $i$-th bin. 

The Fisher matrix for angular power spectra is given by
\begin{align}
	\F_{\alpha\beta}^{\rm 2pt} = \sum_{\X\Y}\sum_{\X'\Y'}\sum_\ell \frac{\partial C_\ell^{\X\Y}}{\partial \lambda_\alpha}({\sf C}^{-1})_\ell^{\X\Y,\X'\Y'}\frac{\partial C_\ell^{\X'\Y'}}{\partial \lambda_\beta}\,\label{F2pt}
\end{align}
We often also include the CMB temperature and polarization data, which we assume to be an independent source of statistical information. 
This is done by allowing the components to run over $\X,\Y\in\{T,E\}$, with $N_\ell^{T,E}$ given by the instrumental noise spectrua for the CMB temperature $T$ and $E$-mode polarization measurements. For Gaussian beams, the noise spectra are typically modeled as
\begin{equation}
	N_\ell^{T,E} = \Delta^2_{T,E}\hs e^{\ell(\ell+1)/\ell_b^2}\, ,
\end{equation}
where $\Delta_{T,E}$ is the detector noise level and $\ell_b\equiv \sqrt{8\log 2}/\theta_b$, with $\theta_b$ the full width at half maximum of the beam. 

\subsubsection{Bispectrum}\label{sec4:bis}

As with the power spectrum case, we count cross-bispectra based on the amount of the overlap between tomographic bins. For the LSST photometric window functions with 16 bins, we include any $(i_1,i_2,i_3)$-bin spectra that satisfy $|i_a-i_b|\le 2$ for $a,b\in\{1,2,3\}$. In this case, the number of $\delta_g\delta_g\delta_g$, $\delta_g\delta_g\kappa$, $\delta_g\kappa\kappa$, and $\kappa\kappa\kappa$ bispectra are $6N_{z}-8$, $3N_{z}-3$, $N_{z}$, and $1$, respectively, giving a total of $10(N_{z}-1)=150$ three-point observables to consider.

The Fisher matrix for angular bispectra is given by~\cite{Babich:2004yc, Yadav:2007rk} 
\begin{align}\label{F3pt}
	\F_{\alpha\beta}^{\rm 3pt} = \frac{f_{\rm sky}}{6}\sum_{\cal XYZ}\sum_{\cal X'Y'Z'}\sum_{\ell_1\ell_2\ell_3}g_{\ell_1\ell_2\ell_3}^2 \frac{\partial B_{\ell_1\ell_2\ell_3}^{\X\Y\Z}}{\partial \lambda_\alpha} (D^{-1})^{\X\X'}_{\ell_1}(D^{-1})^{\Y\Y'}_{\ell_2}(D^{-1})_{\ell_3}^{\Z\Z'}\frac{\partial B_{\ell_1\ell_2\ell_3}^{\cal X'Y'Z'}}{\partial \lambda_\beta}\, ,
\end{align}
where $\X,\Y,\Z\in\O$, the geometric factor $g_{\ell_1\ell_2\ell_3}$ was defined in \eqref{YlmYlm}, 
and $(D^{-1})^{\X\X'}_\ell$ is the inverse of $D^{\X\X'}_\ell\equiv C^{\X\X'}_\ell+\delta_{\X\X'}N^{\X}_\ell$. 
Rotational invariance and parity imply that we only get non-vanishing contributions when the sum of three multipoles is an even number, i.e.~when~$\frac{1}{2}(\ell_1+\ell_2+\ell_3)\in\mathbb{Z}$.\footnote{When including parity-odd observables such as the $B$-mode polarization, it is possible to construct parity-conserving bispectra for odd $\frac{1}{2}(\ell_1+\ell_2+\ell_3)$~\cite{Meerburg:2016ecv, Duivenvoorden:2019ses}.} This allows us to restrict the sum over multipoles to configurations with $\ell_{1}\geq\ell_{2}\geq\ell_{3}$, multiplied by appropriate symmetry factors.

To facilitate the computation of \eqref{F3pt}, we choose to sample the bispectra with a linear bin-width $\Delta\ell=8$, with $\ell_{\rm min}=20$.  We then approximate the Fisher matrix by the rescaling
\begin{equation}
\F_{\alpha\beta}^{\rm 3pt} \approx \frac{n_{\rm total}}{n_{\rm sample}}\, \F_{\alpha\beta,\, \text{sample}}^{\rm 3pt}\, ,
\end{equation} 
where $n_{\rm total}$ is the total number of possible multipole configurations, $n_{\rm sample}$ is the number of the non-vanishing sampled configurations, and $\F_{\alpha\beta,\, \text{sample}}^{\rm 3pt}$ is the Fisher matrix obtained by summing over the sampled bispectra. 
We make a further simplification by neglecting the galaxy cross-bispectra in the derivative part of the Fisher matrix while keeping them in the covariance matrix. 
Details on why this is reasonable will be given in \S\ref{sec:crossbin}.

\subsection[Forecast Results for CMB$\times$LSS]{Forecast Results for CMB$\boldsymbol{\times}$LSS}\label{sec:res}

In this section, we present the results of our forecast on cosmological parameters using the Fisher formalism just described. Here we focus on the combination of two forthcoming experiments: CMB-S4~\cite{Abazajian:2016yjj,Abazajian:2019eic} and LSST~\cite{Abell:2009aa}. 
The results for other galaxy surveys as well as a summary of the experimental specifications are given in Appendix~\ref{app:more}.

We compute angular spectra using a hybrid approach: we use FFTLog for low $\ell$ (for $\ell<k_{\rm L}\chi(\bar z_i)$ with $k_{\rm L}=0.05\hMpc$ and $\bar z_i$ the mean redshift of the $i$-th bin) and the Limber approximation for high $\ell$ up to $\ell_{\rm max}(z) = k_{\rm max}\chi(\bar z_i)$. The one-loop corrections to power spectra markedly contribute only at high $\ell$ and are therefore computed with the Limber approximation. 
We do not separate the baryon acoustic oscillation (BAO) from the broadband shape, but instead vary the whole power spectra and bispectra in our forecast.
We choose $k_{\rm max}=0.1\hMpc$ and $k_{\rm max}=0.3\hMpc$ for tree and one-loop spectra, respectively, based on the fact that the SPT framework starts to break down at $k\simeq 0.1\hMpc$ and $z=0$~\cite{Lazanu:2015rta}, while one-loop predictions of EFTofLSS agree well with simulations up to $k\simeq 0.3\hMpc$~\cite{Carrasco:2013mua,Baldauf:2015aha,Philcox:2020vvt}. We use the same $k_{\rm max}$ for the entire redshift range that we consider, which is a conservative choice given that perturbations are less nonlinear at higher redshifts.

\begin{table}[t!]
\centering
\begin{tabular}{c c c c c c c}
\toprule
 & \multicolumn{2}{c}{LSST} & \multicolumn{2}{c}{$\times$ S4 Lensing} & \multicolumn{2}{c}{+ S4 T\&P} \\
 	\cmidrule(lr){2-3} \cmidrule(lr){4-5}\cmidrule(lr){6-7}
  & PS & PS+B &  PS & PS+B & PS & PS+B\\
\midrule 
$\sigma(H_0)$ [km/s/Mpc] & 4.35 & 2.24 & 1.71 & 1.04 & 0.31 & 0.16\\
$10^5\hs\sigma(\omega_b)$ & 498 & 256 & 196 & 115 & 2.5 & 2.3\\
$10^4\hs\sigma(\omega_c)$ & 190 & 94 & 57 & 35 & 3.5 & 2.0\\
$10^{9}\hs\sigma(A_s)$& 0.54 & 0.28 & 0.079 & 0.036 & 0.0050 & 0.0043\\
$10^4\hs\sigma(n_s)$ & 537 & 221 & 72 & 52 & 15 & 12\\
\midrule
$\sigma(M_{\nu})$ [meV] & 284 & 99 & 102 & 42 & 21 & 12\\
$\sigma(\fnl)$ & 3.46 & 1.77 & 2.01 & 1.02 & 1.85 & 0.98\\

\bottomrule
\end{tabular} 
\caption{Forecasted 1$\sigma$ marginalized errors on the parameters of the reference ${\nu\Lambda\text{CDM}+\fnl}$ cosmology for LSST combined with CMB-S4. We use  $C^{gg}, B^{ggg}$ for the constraints from LSST and unlensed $C^{TT},C^{TE},C^{EE}$ for CMB temperature and polarization (T\&P). For combinations between two experiments, we also include cross-correlations between galaxies and CMB lensing: $C^{g\kappa},C^{\kappa\kappa}, B^{gg\kappa},B^{g\kappa\kappa},B^{\kappa\kappa\kappa}$. ``PS'' and ``B'' refer to the one-loop power spectra with $k_{\text{max}}=0.3\hMpc$ and the tree bispectra with $k_{\text{max}}=0.1\hMpc$, respectively. }
\label{tab:LSSTresults}
\end{table} 

\begin{figure}[t!]
    \centering
         \includegraphics[scale=.82]{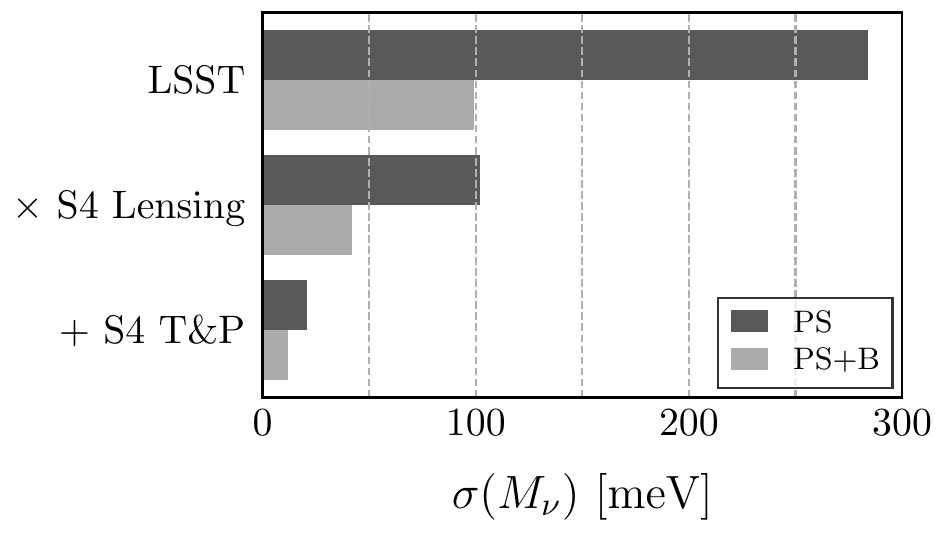}
 \hskip -5pt \includegraphics[scale=.82]{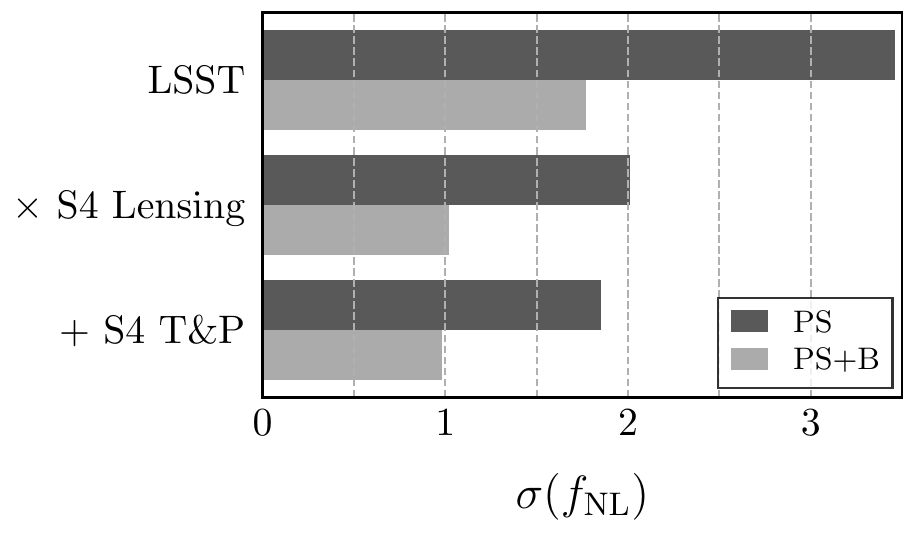} \\
    \caption{Forecasted 1$\sigma$ marginalized errors on $M_\nu$ and $\fnl$ for LSST combined with CMB-S4 (same as the bottom two rows of Tab.~\ref{tab:LSSTresults}). The darker bars show the constraints from the one-loop power spectra and the lighter bars show the constraints with the bispectra also included.}
    \label{fig:barhMnufnl}
\end{figure}

Table~\ref{tab:LSSTresults} presents the forecasted 1$\sigma$ constraints from CMB-S4 and LSST on parameters of the reference ${\nu\Lambda\text{CDM}+\fnl}$ cosmology (see also Fig.~\ref{fig:barhMnufnl} for a more direct visualization on the constraints on $M_\nu$ and $f_{\rm NL}$). 
Our result shows that it is possible to reach $\sigma(M_\nu)=21\,$meV from the power spectrum information alone and $\sigma(M_\nu)=12\,$meV when further adding the bispectrum information. 
Since the inverted hierarchy of neutrino masses requires $M_\nu\ge 100\,$meV, we see that the combination of these two experiments will be capable of ruling out the inverted hierarchy at about $4\sigma$ confidence if $M_\nu\simeq60\,$meV. 
It is worth mentioning that these constraints do not assume any prior information on the optical depth. Our results should be compared to that of~\cite{Yu:2018tem}, which showed that the cross-spectra between galaxies from LSST and CMB lensing from CMB-S4 can reach $\sigma(M_\nu)=68\,$meV without any prior information on the optical depth. (To be more precise, this corresponds to the LSST Gold galaxy sample with non-overlapping tomographic bins and tree-level power spectra with $k_{\rm max}=0.1\hMpc$.) 
We improve upon this result by a factor of 5, which mainly comes from both adding the one-loop corrections to power spectra, which enables us to extend $k_{\rm max}$ to $0.3\hMpc$ from $0.1\hMpc$, and adding the bispectrum information. 
Also, we see that the constraints on the local non-Gaussianity amplitude $\sigma(\fnl)\simeq 1$ will be achievable, which is about a factor of 5 improvement over the current constraint from Planck~\cite{Akrami:2019izv}.

\begin{figure}[ht!]
    \centering
         \includegraphics[width=\textwidth]{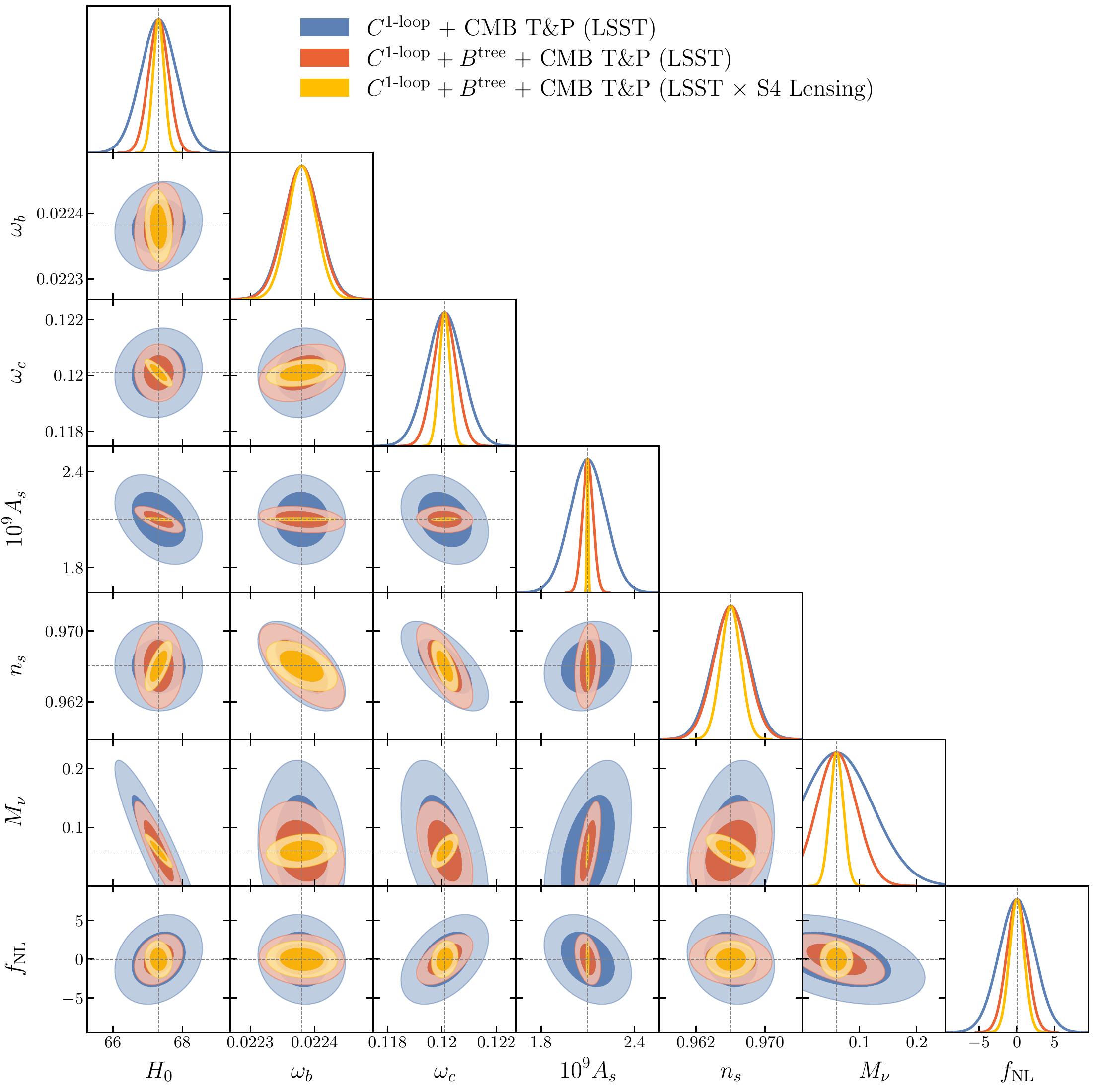}
    \caption{Forecasted 1$\sigma$ and 2$\sigma$ constraints on the parameters of the reference ${\nu\Lambda\text{CDM}+\fnl}$ cosmology for LSST combined with CMB-S4. The blue contours represent the constraints from the galaxy one-loop power spectrum, and the yellow contours represent the constraints including the galaxy tree bispectrum. The red contours also include the cross-power spectra and cross-bispectra between galaxies and CMB lensing. All three cases include the CMB temperature and polarization power spectra from CMB-S4. We use $k_{\rm max}=0.1 \hMpc$ for tree and $k_{\rm max}=0.3 \hMpc$ for one-loop spectra. The dashed lines indicate the fiducial values of parameters given in Tab.~\ref{tab:param}. ($H_0$ is in units of km/s/Mpc and $M_\nu$ is in units of eV.)}
    \label{fig:contour}
\end{figure}

Figure~\ref{fig:contour} shows the $1\sigma$ and $2\sigma$ contours, showing correlations between different parameters. We see that when adding bispectra, the parameter constraints improve by a factor of $3$ for $M_\nu$ and a factor of 2 for the rest the parameters. In turn, adding galaxy and CMB lensing cross-correlations improves the results by twofold for most parameters, except for $n_s$ and $A_s$ whose constraints improve by factors of 4 and 8, respectively.
Because including CMB lensing strongly breaks the degeneracy between $A_s$ and the linear bias $b_\delta$, it significantly improves the constraint on $A_s$. Similar results can also be found in~\cite{Schmittfull:2017ffw}, in which they show that the precision of $\sigma_{8}$ improves by more than a factor of 10 with cross-correlations of CMB lensing. 

\subsubsection{CMB Lensing} 

CMB lensing serves as a tracer for the underlying matter density field. By cross-correlating it with a galaxy survey, we can circumvent the cosmic variance limited by the survey volume, and also partially cancel the degeneracy between different cosmological parameters. The idea behind this is simply that by making use of two different tracers, we can partially cancel the random processes that form the same underlying density field~\cite{Seljak:2008xr}. For this to occur, it is important that the two experiments observe the same patch of the sky, so that their measured spectra are correlated.

As a consequence, the parameter constraints show a marked improvement with the addition of cross-correlations between CMB lensing and galaxies. 
First, it is well known that $g\kappa$ cross-correlations can help breaking the degeneracy between the amplitude parameters $A_s$ and $b_\delta$ of the tree galaxy power spectrum, since $C^{gg}\propto b_\delta^2 A_s $ and $C^{g\kappa}\propto b_\delta A_s $. Indeed, we see that adding CMB lensing cross-correlations has a significant effect on measurements of $A_s$: we obtain an improvement factor 7 and 8 on its constraint, when adding the cross power spectra and bispectra, respectively. 
In addition, we obtain a factor of 3 and 2 improvement in the constraint on $M_{\nu}$ upon inclusion of the cross power spectra and bispectra, respectively, compared to galaxy-only correlations. 
A slightly less, about 50\% error reduction is instead seen for the $\fnl$ constraint. This is partly because galaxy-only spectra, being proportional to higher powers of the scale-dependent galaxy bias $\delta_g$ than cross-spectra, have a stronger dependence on $\fnl$.

\subsubsection{One-Loop Power Spectrum} 

\begin{figure}[ht!]
    \centering
         \includegraphics[width=\textwidth]{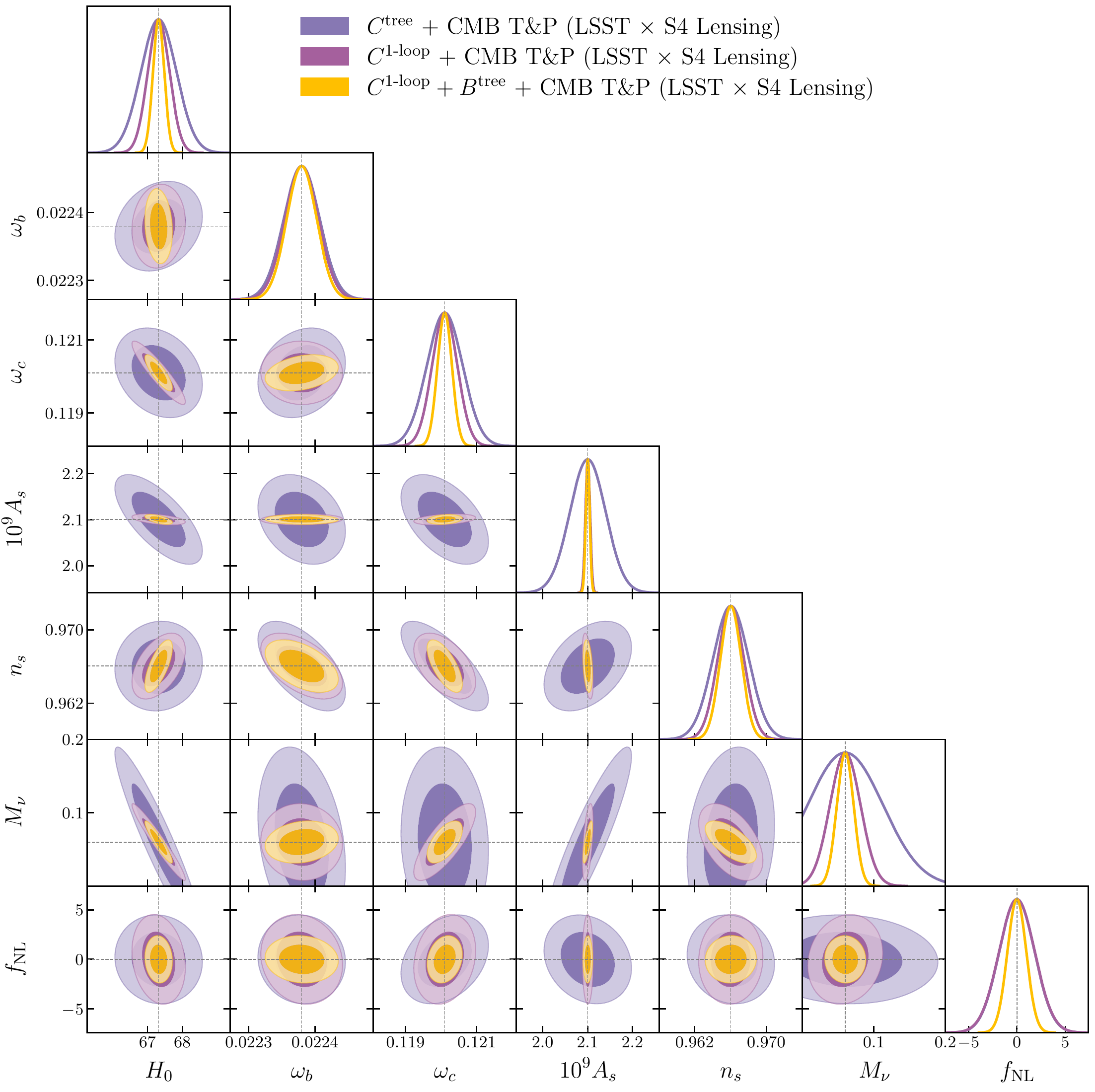}
    \caption{Forecasted 1$\sigma$ and 2$\sigma$ constraints on the parameters of the reference ${\nu\Lambda\text{CDM}+\fnl}$ cosmology for LSST combined with CMB-S4. The two types of purple contours compare constraints between tree and one-loop power spectra, whereas the yellow contours show the constraints from one-loop power spectra combined with tree bispectra. All three cases include the CMB temperature and polarization power spectra from CMB-S4 as well as cross-correlations between galaxies and CMB lensing. We use $k_{\rm max}=0.1 \hMpc$ for tree and $k_{\rm max}=0.3 \hMpc$ for one-loop spectra. The dashed lines indicate the fiducial values of parameters given in Tab.~\ref{tab:param}. ($H_0$ is in units of km/s/Mpc and $M_\nu$ is in units of eV.)}
    \label{fig:contour2}
\end{figure}

Adding the one-loop corrections to the matter power spectrum generally has two competing effects.
First, they introduce 4 new nuisance bias parameters, which, when marginalized over, result in degradation of overall constraints. 
At the same time, however, the new shape dependences they introduce can help breaking the degeneracies between certain parameters. For example, the linear bias and the scalar amplitude are fully degenerate at tree level because $P^{gg}_{\rm tree}\propto b_\delta^2 A_{s}$, whereas at one loop there are additional terms proportional to $b_\delta^2 A_{s}^{2}$.
Moreover, extending the theory to one loop allows us to increase the maximum wavelength $k_{\rm max}$ up to which we can trust perturbation calculations, allowing access to an increased number of useful Fourier modes.  
The combined effect then generally depends on the value of $k_{\text{max}}$. In our analysis, we set $k_{\rm max}=0.1\hMpc$ for tree power spectra and $k_{\rm max}=0.3\hMpc$ for one-loop power spectra. With this choice, we find that the loop corrections improve the constraints roughly by a factor of $2.5$ for $M_{\nu}$, a factor of $8$ for $A_s$, and a factor between 1--2 for the rest of the parameters, as shown in Fig.~\ref{fig:contour2}. 
	
A similar but stronger conclusion was reached in~\cite{Baldauf:2016sjb} regarding the impact of adding the one-loop galaxy power spectrum, which showed about a factor of 5 improvement on the neutrino mass constraint for a forecast conducted in momentum space. Although the different modeling makes a direct comparison difficult, in general it is expected that we gain less number of quasi-nonlinear modes in angular space as we increase $k_{\rm max}$, since these modes are restricted to the 2D projected surfaces perpendicular to the line of sight. To recover the full 3D information, it is necessary to simultaneously reduce the redshift uncertainty $\Delta z$ to have roughly the same size as the smallest wavelength we can probe, given by $k_{\rm max}$~\cite{Asorey:2012rd}. We have used the same tomographic binning scheme for both tree and one-loop power spectra in our comparison for simplicity, which accounts for the less dramatic impact of one-loop power spectrum that we have found. 
		
\subsubsection{Bispectra} 
At tree level, the matter bispectrum does not induce new bias parameters when added to the one-loop power spectrum, and its inclusion therefore helps to further break parameter degeneracies. 
Figure~\ref{fig:contour2} shows a visualization of the improved parameter constraints after adding the tree bispectra of galaxies and CMB lensing.
We see that $\omega_b$, $A_s$ and $n_s$ are already well-constrained by the one-loop power spectra and CMB T\&P, and the bispectra do not add much more information. 
In contrast, the constraints on $H_0$, $\omega_c$, $M_\nu$ and $\fnl$ display a notable improvement with the addition of the bispectra, with their errors reduced by a factor of 1.5--2.5.

Let us also comment on other recent works that studied the impact of adding bispectra on parameter constraints. 
For example, \cite{Chudaykin:2019ock} presented forecasted parameter constraints from the galaxy one-loop power spectrum and tree bispectrum for Euclid. Although they used a different, Markov Chain Monte Carlo (MCMC) method for their forecast and considered only galaxy spectra in Fourier space, they found an almost factor of 2 improvement on the constraint on $M_\nu$ when adding the bispectrum information. 
Specifically, they found the neutrino mass constraint for $P^{gg}_{\text{1-loop}}$ combined with Planck went down from $\sigma(M_\nu)=23\,$meV to $11\,$meV when further including the tree bispectrum.
In comparison, our results show $\sigma(M_\nu)=21\,$meV from $C^{gg,g\kappa,\kappa\kappa}_{\text{1-loop}}$ combined with CMB-S4 and $\sigma(M_\nu)=12\,$meV after the bispectrum information has been added. 
In another interesting work~\cite{Hahn:2019zob}, the authors used $N$-body simulations to study the effects of nonlinear clustering on parameter constraints. For $k_{\rm max}=0.5\hMpc$, they showed that the constraint on $M_{\nu}$ (with a Planck prior) including the bispectrum is about 1.8 times tighter than that of the power spectrum alone. 
Further, an MCMC analysis of \cite{MoradinezhadDizgah:2020whw} showed that the constraint on local $\fnl$ significantly improves with the addition of bispectra.
Despite the number of differences in theoretical modeling, all of these works clearly indicate that the bispectrum contains a significant amount of information in addition to the power spectrum, which will be useful for tightening parameter constraints, especially that of $M_\nu$ and $\fnl$, in near-future surveys.

\subsubsection{Limber vs.~FFTLog}  

Using the Limber approximation has a number of ramifications in parameter estimation. 
First, as was indicated in Fig.~\ref{fig:ClComp} and Fig.~\ref{fig:BlllComp}, the Limber approximation has a tendency to under-predict the signal at large scales, therefore leading to larger errors on parameter constraints in general. 
But more importantly, an incorrect modeling of angular correlation functions due to the Limber approximation can lead to systematic biases in parameters' best-fit values from their {\it true} values. 
While these errors have been sufficiently small so far for experiments that probed small angular scales, it is no longer guaranteed to so for the next generation of surveys. 

In a typical cosmological analysis, we extract the best-fit values of parameters by maximizing the likelihood for an observable in a given theoretical model. If a wrong theoretical model is assumed, then inferred parameters will be displaced from their values in the true underlying cosmology. Assuming a Gaussian likelihood, the linear displacement of wrongly-inferred parameters from their true values is estimated by~\cite{Bernal:2020pwq}
\begin{align}
	\Delta\lambda_{\alpha} &= \sum_{\beta}(\F^{2\rm pt}_{\text{Limber}})^{-1}_{\alpha\beta}	\left[\sum_{\X\Y}\sum_{\X'\Y'}\sum_{\ell}\left(C_{\ell}^{\X\Y}-C_{\ell,\hs\text{Limber}}^{\X\Y}\right)\left({\sf C}^{-1}_{\text{Limber}}\right)_{\ell}^{\X\Y,\X'\Y'}\frac{\partial C_{\ell,\hs\text{Limber}}^{\X'\Y'}}{\partial\lambda_{\beta}}\right],
\end{align}
where $C_\ell^{\X\Y}$ is the true underlying power spectrum and the subscript `Limber' means that the quantity is computed under the Limber approximation. 

\begin{figure}[t!]
    \centering
         \includegraphics[scale=.68]{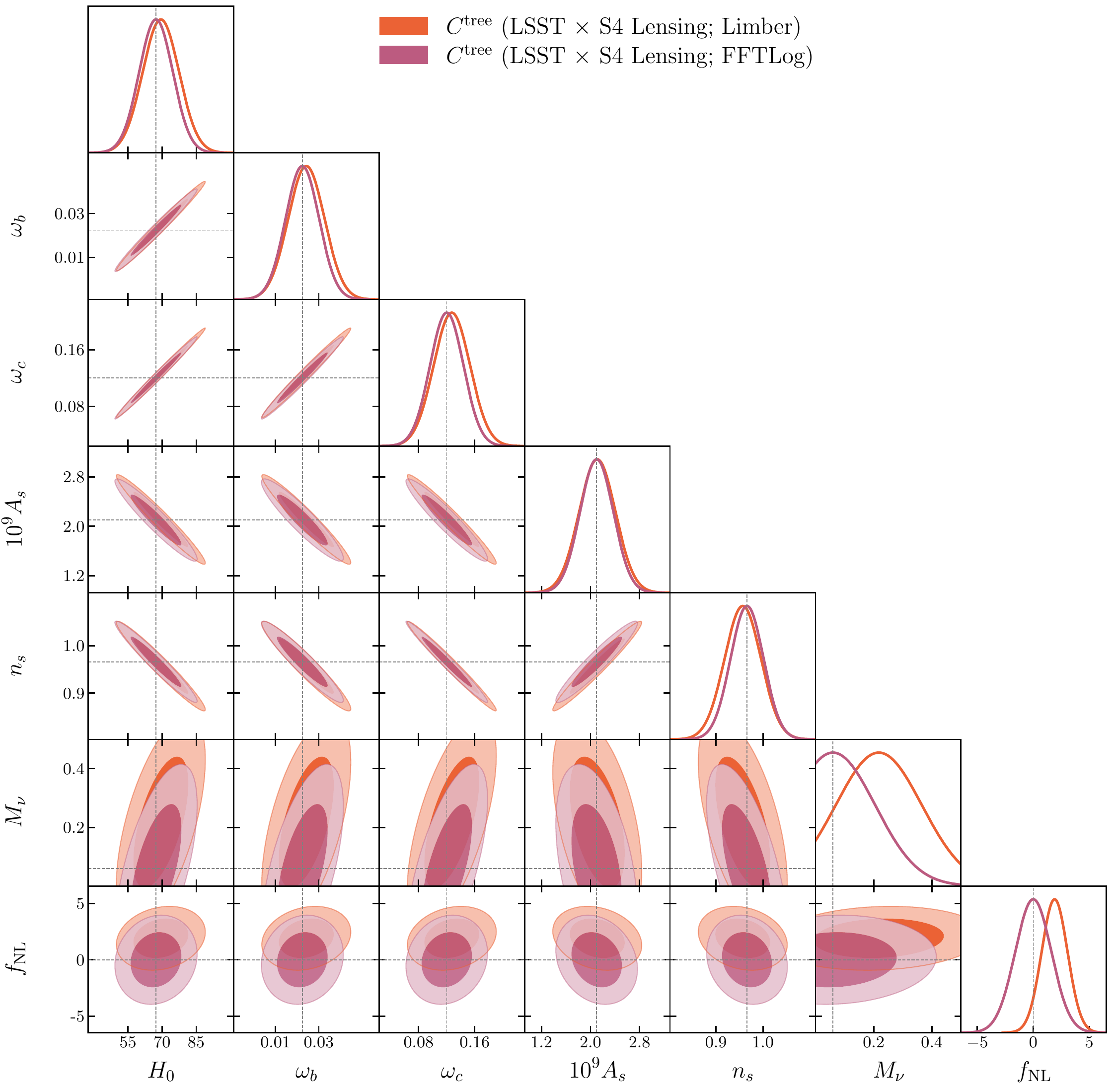}
    \caption{Forecasted $1\sigma$ and $2\sigma$ constraints on the parameters of the reference ${\nu\Lambda\text{CDM}+\fnl}$ cosmology for LSST combined with CMB-S4. The red contours show the constraints obtained using the Limber approximation, while the purple contours show the constraints obtained using the FFTLog method. The results here are obtained from the tree power spectra for LSST combined with CMB-S4 lensing, split into 6 tomographic bins. We use $k_{\rm max}=0.1\hMpc$ and $\ell_{\rm min}=10$. The dashed lines indicate the fiducial values of parameters given in Tab.~\ref{tab:param} ($H_0$ is in units of km/s/Mpc and $M_\nu$ is in units of eV.) }
    \label{fig:limberfftlog10}
\end{figure}

To understand the impact of the Limber approximation more concretely, we  also performed a simpler forecast for a wider multipole range with $\ell_{\rm min}=10$, using only tree power spectra (from LSST galaxies cross-correlated with CMB-S4 lensing) for 6 redshift bins with edges given by the set $\{0,0.5,1,2,3,4,7\}$.
\footnote{We used $\ell_{\text{min}}=20$ for our main forecast in Tab.~\ref{tab:LSSTresults}, corresponding to a maximum angular separation of $\theta_{\rm max}\simeq 9^\circ$. This rather conservative choice of $\ell_{\text{min}}$, however, resulted in not-so-large differences between FFTLog and the Limber approximation, giving about $5\%$ differences for most forecasted constraints.} 
The result is visualized in Fig.~\ref{fig:limberfftlog10}, which shows a comparison of parameter constraints from the tree power spectra computed with FFTLog and the Limber approximation. We can see that Limber's method induces biases on the best-fit values of most parameters, which are especially prominent for $\fnl$ and $M_{\nu}$ that show 1--2$\sigma$ shifts. The physical origin of the shift for $\fnl$ is easily understood: the galaxy power spectrum constrains $\fnl$ through the scale-dependent bias \eqref{scaledependentbias}, which only affects very large scales. Similarly, the shift for $M_\nu$ arises due to the fact that the multipole corresponding to the neutrino free-streaming scale $k_{\rm fs}$ for our fiducial cosmology is smaller than the scale above which it is safe to use the Limber approximation; for example, $k_{\rm fs}\lesssim 0.02\hMpc$ for $M_\nu=60\,$meV for the relevant redshifts (c.f.~\eqref{kfs}), whereas typically the Limber approximation works well for $k\gtrsim 0.05\hMpc$. For this reason, we also expect this shift to change with other fiducial cosmologies. 
Refs.~\cite{Bellomo:2020pnw, Bernal:2020pwq} also showed that the Limber approximation can induce a large systematic bias in $\fnl$ constraints for future galaxy surveys.\footnote{Refs.~\cite{Bellomo:2020pnw, Bernal:2020pwq} studied biases from the galaxy-only power spectrum and worked with a lower $\ell_{\rm min}$, which led to a much larger, potentially greater than $10\sigma$ shift for $\fnl$. In addition, they found that not accounting for lensing magnification \eqref{magnification} may also result in non-negligible biases. Our findings show that these biases can persist at the $2\sigma$ level, even after combining with less-Limber-sensitive CMB lensing power spectra (see Fig.~\ref{fig:ClComp}) and using a relatively larger $\ell_{\rm min}=10$, and that the constraints on $M_\nu$ suffer from similar biases.}

Over the next few years, the most likely source of information on cross-correlations of galaxies with CMB lensing is going to be Dark Energy Spectroscopic Instrument (DESI) for galaxies, and Planck and Atacama Cosmology Telescope (ACT)~\cite{Louis:2014aua} for the CMB. 
A short comparison of the forecast results between DESI and LSST is provided in Appendix~\ref{app:more}. 
Since these experiments access relatively higher $\ell$ modes (with $\ell_{\rm min}\simeq 100$), Limber-based calculations are likely to be sufficient for current datasets.
More futuristically, we will have data from spectroscopic surveys such as Euclid~\cite{Amendola:2016saw} and the Roman Telescope~\cite{Spergel:2015sza} in the coming decade. These will measure galaxy redshifts at higher precision than LSST, which would result in narrower tomographic bins and, as a consequence, make the Limber approximation worse. It would be instructive to scrutinize the applicability of the Limber approximation for a wider set of experiments than the ones considered in this work.

\subsubsection{Redshift Space Distortion} 

The effects of RSD on angular galaxy spectra strongly depend on the type of the window function used. In general, as the window function becomes wider, the peculiar velocities of galaxies are averaged out more within a selected tomographic bin.  For the photometric LSST window function with $\Delta z\approx 0.1$, the RSD contribution leads to about $10$--$20$\% enhanced signal of the galaxy auto-spectra in the range $10\lesssim \ell\lesssim 50$, which in turn leads to small percent-level changes in forecasted errors of most parameters. 
It turns out that the RSD effects on cross-bin spectra are more significant and cannot be neglected even for high multipoles if the two bins do not overlap; however, their small overall amplitudes relative to auto-spectra meant that they did not leave a significant impact on our parameter forecast either.
In general, the Limber approximation and the RSD cannot be made compatible with each other; that is, the RSD only becomes relevant when the Limber approximation breaks down due to narrow tomographic bins (see~\cite{DiDio:2018unb} for more discussions). 

\subsubsection{Bias Parameters} 

\begin{figure}[t!]
    \centering
    \includegraphics[height=5cm]{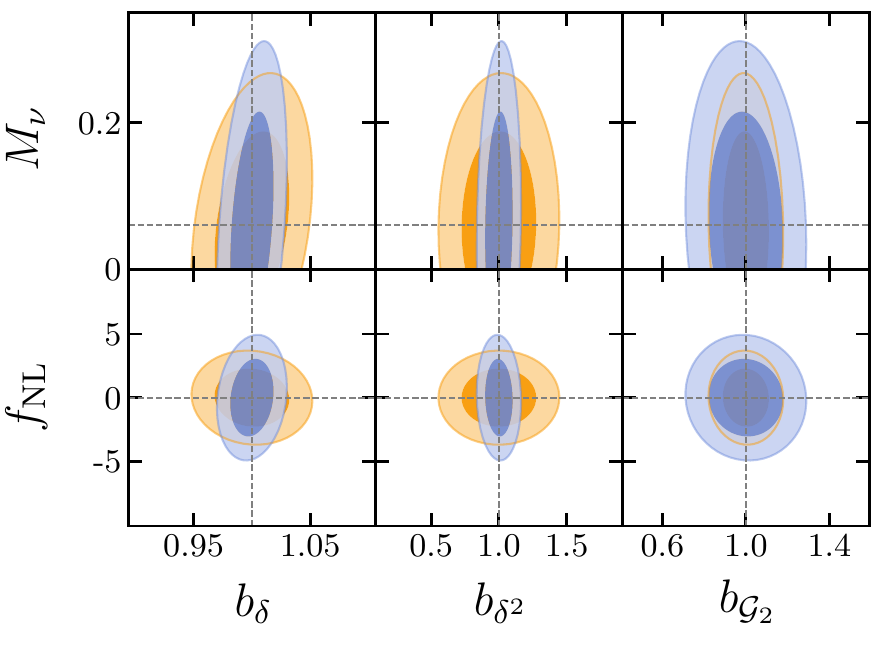}
\quad    \includegraphics[height=5cm]{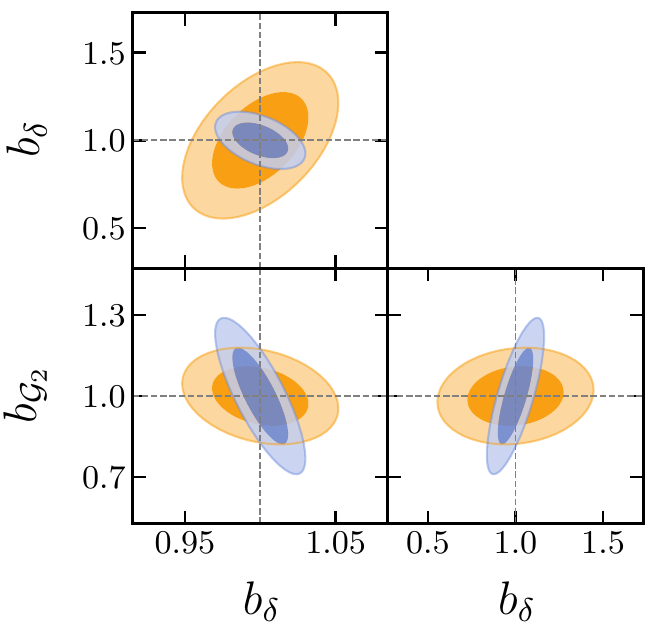}
    \caption{Forecasted constraints on $M_\nu$, $f_{\rm NL}$, and the bias parameters in a single tomographic bin $z\in[0.4,0.6]$. 
    The blue and yellow contours show the constraints from $C^{\text{1-loop}}$ and $B^{\rm tree}$ (LSST combined with CMB-S4 Lensing), respectively.}
    \label{fig:contour_bias_lowz}
\end{figure}

Including one-loop power spectra and tree bispectra allows us to break most of the degeneracies between biases and cosmological parameters, e.g.~that between $b_\delta$ and $A_s$, as we explained earlier. 
As an illustration, we show in Fig.~\ref{fig:contour_bias_lowz} the forecasted constraints on $M_\nu$, $f_{\rm NL}$, together with the bias parameters and the degeneracies among them, for simplicity in a single tomographic bin. 
First, we clearly see that $M_{\nu}$ and $f_{\rm NL}$ are nearly uncorrelated with the bias parameters. This is because they have very different shape contributions to the power spectrum: $\fnl$ ($M_\nu$) leads to a scale-dependent growth (suppression) at low (intermediate) $\ell$, while the nonlinear biases modify the shape at high $\ell$. Since this behavior is true for any redshift, they remain uncorrelated in other tomographic bins, independent of the particular bias model used. (We adopt a specific co-evolution bias model described in Appendix~\ref{app:bias}.) 
On the other hand, we see that the degeneracies between the bias amplitudes themselves are not fully broken (within a single tomographic bin). 
These residual degeneracies could be broken, for example, with the addition of the tree galaxy trispectrum~\cite{Lee:2020ebj}, since it depends on the same set of nonlinear biases but does not itself introduce extra nuisance parameters. 

\subsubsection{Tomographic Cross-Spectra}\label{sec:crossbin}

When photometric errors are taken into account, there is a moderate amount of overlap between adjacent tomographic bins. To see this, let us refer back to Fig.~\ref{fig:ClComp} that shows the galaxy auto- and cross-spectrum for two overlapping Gaussian window functions centered at $z=1,1.2$, both with $\sigma_z=0.1$. We see that the cross-spectrum has an amplitude that is as large as about 20\% relative to the auto-spectrum. 
The two auto-spectra at $z=1,1.2$ therefore cannot be treated as fully independent observables, and their cross-spectrum should be included to account for the degenerate information between them. 

Having said that, computing all possible cross-bin correlations in a Fisher forecast can nonetheless be quite time-consuming, especially for bispectra. As we mentioned in \S\ref{sec4:bis}, we have thus chosen to neglect the derivative part of the cross-bin bispectra in the main forecast. 
To check the validity of this approximation, we divided the 16 tomographic bins into 4 smaller subsets, and then compared the forecasted constraints with and without the cross-bin correlations in the derivative part. This resulted in at most $10\%$ difference in the forecasted constraints on all cosmological parameters of the $\nu\Lambda$CDM+$\fnl$ cosmology between the two results. 
Although the parameter constraints were not substantially affected by this assumption, it would nevertheless be desirable to include the full contribution from cross-bin bispectra in a future analysis with a more advanced computational method.

The contribution from cross-bin correlations becomes more relevant when there are larger overlaps between different tomographic bins. The situation can be contrasted with galaxy cross-spectra from non-overlapping window functions, which have smaller amplitudes and even identically vanish under the Limber approximation. 
We find that fully incorporating the cross-bin spectra from the LSST photometric window functions with $\sigma_z= 0.05(1+z)$ (see Appendix~\ref{app:more} for a precise definition) leads to about 25\% degradation in the constraint on $M_\nu$ compared to the case of non-overlapping window functions considered in~\cite{Schmittfull:2017ffw, Yu:2018tem}.

\subsubsection{CMB Prior} 

We add the information from the unlensed CMB temperature and polarization (T\&P) spectra from CMB-S4 in most of our results, without additional priors on any of the parameters considered.
Doing so, we find that the constraint on $M_{\nu}$ from the galaxy \& CMB lensing power spectra and that also including the bispectra become 5 and 3.5 times better, respectively. (The improvement factors for other parameters vary between 5 and 80, except for $\fnl$ for which there is less than 10\% difference.)
In comparison, the authors of \cite{Yu:2018tem} obtained roughly a twofold improvement on the $M_\nu$ constraint after adding CMB T\&P to LSST cross-correlated with CMB lensing. 
This difference may be understood as coming from the one-loop corrections to power spectra. Even though the one-loop power spectra have more nuisance parameters due to higher-order biases, the degeneracies between cosmological parameters actually become smaller with the addition of the one-loop corrections. Note that this information is not visible from marginalized 1D constraints. The CMB T\&P information is then capable of breaking the degeneracies between the higher-order biases, and adding it therefore has a larger impact on the one-loop power spectra.

\begin{tcolorbox}[title=Summary of new results]
\begin{itemize}
	\item 
	While the Limber approximation leads to small differences in the sizes of marginalized errors compared to the more accurate FFTLog, it can introduce systematic biases in the best-fit values of cosmological parameters. For example, we find 1--2$\sigma$ biases for $M_{\nu}$ and $\fnl$ at tree level for $\ell_{\rm min}=10$. 
	\item
	Adding the one-loop power spectra, tree bispectra, and cross-correlations between galaxies and CMB lensing each improves the overall parameter constraints by a factor of 1.5--2.5 for parameters $H_0$, $\omega_c$, $M_\nu$, and $\fnl$. 
	\item A significant (over $3\sigma$) detection of the total neutrino mass is possible, if we combine the cross-power spectra and cross-bispectra between galaxies and CMB lensing (including the CMB T\&P information). For $M_\nu\simeq 60$~meV, this also rules out the inverted mass hierarchy.
\end{itemize}
\end{tcolorbox}

\section{Conclusions}\label{sec:con}

Cross-correlations of galaxies with CMB lensing have a great potential to improve parameter constraints in future surveys. 
In this paper, we presented a detailed forecast on cosmological parameters from the combination of LSST and CMB-S4, highlighting the constraints on two parameters that extend the vanilla $\Lambda$CDM cosmology: the total neutrino mass $M_\nu$ and the local non-Gaussianity amplitude $\fnl$. 
In particular, we have extended the previous works~\cite{Schmittfull:2017ffw, Yu:2018tem} in a number of directions. 

First, our baseline theoretical model includes angular bispectra at tree level. In Fourier space, it has so far been demonstrated that the inclusion of the galaxy bispectrum can substantially improve parameter constraints~(see e.g.~\cite{Ruggeri:2017dda, Chudaykin:2019ock, Hahn:2019zob} for recent works). However, its angular counterpart as well as its cross-bispectra with CMB lensing are comparatively much less studied, partly due to their computational complexity. In our work, we found that angular bispectra, too, carry substantial statistical information in addition to power spectra, improving overall parameter constraints by two- or threefold. Moreover, including cross-correlations with CMB lensing further improves the constraints by approximately a factor of $2$ for most parameters, except for $n_s$ and $A_s$ whose constraints become $4$ and $8$ times better, respectively. In particular, our findings demonstrate that it would be possible to reach $\sigma(M_\nu)=12\,$meV and $\sigma(\fnl)=1$ with the combination of LSST and CMB-S4. 

In addition, we have included the one-loop corrections to angular power spectra using the effective field theory framework.
The inclusion of these nonlinear corrections has the effect of breaking certain parameter degeneracies---especially the one between $A_s$ and $b_\delta$---and allows us to work with a slightly larger choice of the momentum cutoff $k_{\rm max}$ compared to a tree-level analysis. 
As a result, we found that the one-loop corrections can yield an eightfold improvement in the constraint on $A_s$ and up to a twofold improvement for the rest of parameters, as shown in Fig.~\ref{fig:contour2}. 

We used the FFTLog algorithm~\cite{Assassi:2017lea, Gebhardt:2017chz, Schoneberg:2018fis} to efficiently compute angular power spectra and bispectra in our analysis, and studied the impact of using the Limber approximation in our forecast for LSST. 
For example, for $\ell_{\rm min}=10$ we found that using purely the Limber approximation can result in about $1$--$2\sigma$ shifts in the parameter constraints for $M_\nu$ and $\fnl$ at tree level. This implies that the commonly-used Limber approximation could be a significant source of systematic bias for LSST. 
In particular, an accurate determination of $M_\nu$ without such systematic bias will be crucial in order to correctly interpret the neutrino mass hierarchy from observational data. 

There are many avenues in which our analysis can be further improved. First, it would be informative to perform a similar analysis for spectroscopic surveys such as Euclid or the Roman Telescope. Due to the large photometric redshift uncertainties that need to be smoothed over, the RSD did not play a considerable role in our forecast for LSST. This also implied that the Limber approximation remained valid over a somewhat large range of multipoles, down to $\ell_{\rm min}=20$. In contrast, relatively smaller redshift errors for spectroscopic surveys would invoke an earlier breakdown of the Limber approximation and, simultaneously, a larger impact induced by the RSD. 
Also, including nonlinear effects such as the IR resummation of the BAO~\cite{Senatore:2014via, Senatore:2017pbn, Lewandowski:2018ywf}, the nonlinear Kaiser effect~\cite{Gebhardt:2020imr}, and the LSS theoretical error~\cite{Baldauf:2016sjb} would help refine our forecasted constraints.
Lastly, it would also be interesting to see the impact of angular trispectra~\cite{Lee:2020ebj} and their contribution to the non-Gaussian (super-sample) covariance~\cite{Takada:2013wfa, Lacasa:2017ufk} in parameter constraints. We leave these to future work.

\paragraph{Acknowledgement} 
We thank George Efstathiou, Tanveer Karim, Marcel Schmittfull, and Blake Sherwin for useful discussions. HL and CD acknowledge support from the Department of Energy (DOE) Grant No.~DE-SC0020223. We acknowledge the use of {\tt CAMB}\footnote{\url{https://camb.info}}~\cite{Lewis:1999bs}, {\tt GetDist}\footnote{\url{https://getdist.readthedocs.io/en/latest}}~\cite{Lewis:2019xzd}, and {\tt quicklens}\footnote{\url{https://github.com/dhanson/quicklens}}~\cite{Ade:2015zua}. 

\newpage
\appendix

\section{Bias Model}\label{app:bias}

\begin{figure}[t!]
    \centering
         \includegraphics[scale=.785]{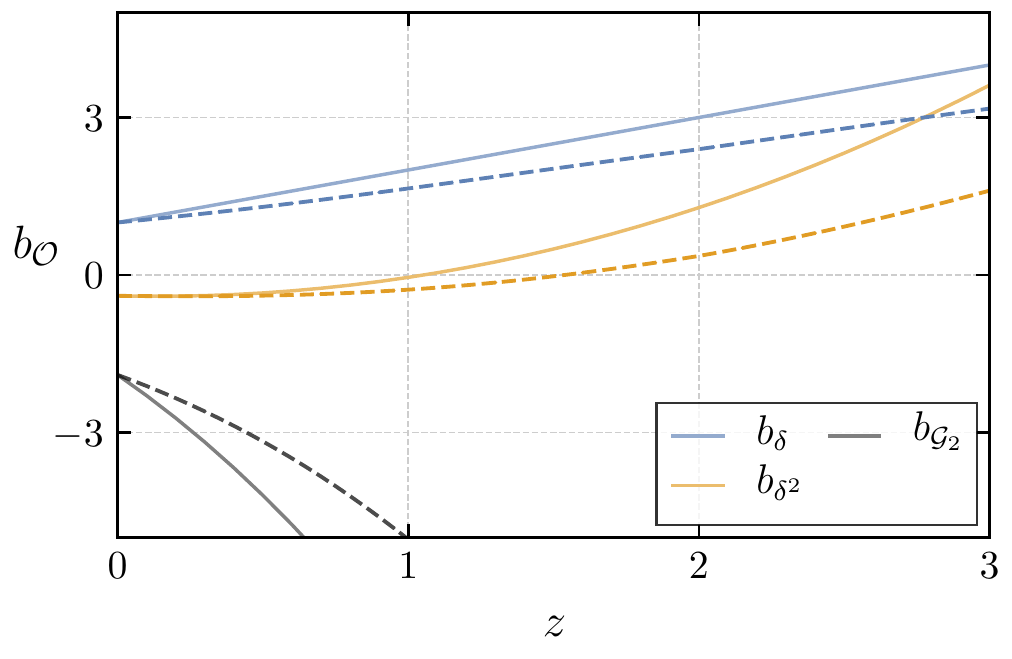}        
          \includegraphics[scale=.785]{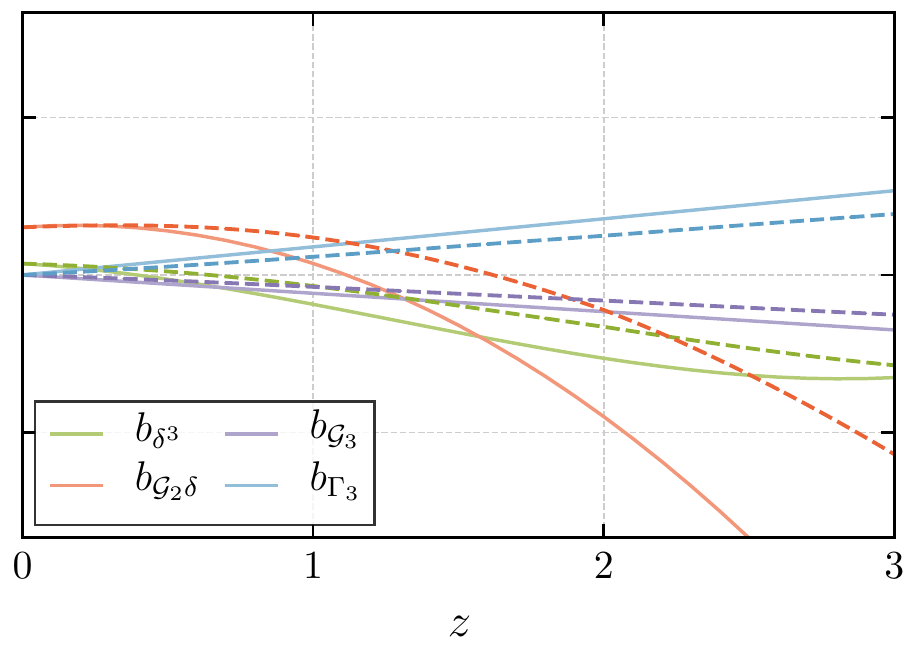}
         \\
    \caption{Bias parameters as functions of redshift $z$ as given in \eqref{eq:bias}, with normalization $\bar b_\O=1$. The left panel shows the biases up to quadratic order, whereas the right panel shows the biases at cubic order. The linear bias has been fixed to $b_1=1+z$ and $b_1=D_+^{-1}(z)$ for the solid and dashed lines, respectively.}
    \label{fig:bias}
\end{figure}

In the local Lagrangian bias model (see~\cite{Desjacques:2016bnm}), the Eulerian biases $b_\delta$, $b_{\delta^2}$, and $b_{\delta^3}$ are given in terms of the corresponding Lagrangian biases as 
\begin{align}
	\begin{bmatrix}
		b_\delta\\ b_{\delta^2}\\b_{\delta^3}
	\end{bmatrix}
	=
	\begin{bmatrix}
		 1 & 0 & 0\\ 
		 \frac{2}{21} & \frac{1}{2} & 0\\
		 0 & -1 &  \frac{1}{6}
	\end{bmatrix}
	\begin{bmatrix}
		b_\delta^L \\ b_{\delta^2}^L \\ b_{\delta^3}^L
	\end{bmatrix}
	+
	\begin{bmatrix}
		1 \\ 0 \\ 0
	\end{bmatrix} .\label{biasset1}
\end{align}
The above matrix can be inverted to express the Lagrangian biases in terms of the Eulerian ones. Using these relations, other Eulerian biases up to cubic order can be expressed in terms of $b_\delta$ and $b_{\delta^2}$ as
\begin{align}
	\begin{bmatrix}
		b_{\G_2}\\b_{\G_2\delta}\\b_{\G_3}\\b_{\Gamma_3} 
	\end{bmatrix}
	=
	\begin{bmatrix}
				\frac{2}{7} & -\frac{2}{7} & \frac{2}{3} \\
				-\frac{16}{147} & \frac{16}{147} & -\frac{8}{7} \\
				\frac{22}{63} & -\frac{22}{63} & 0 \\
				-\frac{23}{43} & \frac{23}{43} & 0\\
	\end{bmatrix}
	\begin{bmatrix}
		1 \\ b_\delta \\ b_{\delta^2}
	\end{bmatrix} .
\end{align}
We use the fitting functions from $N$-body simulations for the biases $b_2=2b_{\delta^2}$ and $b_3=6b_{\delta^3}$ in terms of $b_1$ given by \cite{Lazeyras:2015lgp}
\begin{align}
		b_{2}&=0.412-2.143b_1+0.929b_1^2+0.008b_1^3\, ,\\
		b_{3}&=-1.028+7.646b_1-6.227b_1^2+0.912b_1^3\, .
\end{align}
Parameterizing the overall amplitudes with $\bar b_\O$, we have
\begin{align}\label{eq:bias}
	b_{\delta} &= \bar{b}_{\delta}\times b_1\, ,\\
	b_{\delta^2} &= \bar{b}_{\delta^{2}}(0.206-1.072b_1+0.465b_1^{2}+0.004b_1^{3})\, ,\\
	b_{\delta^{3}} &= \bar{b}_{\delta^{3}}(-0.171+1.274b_1-1.038b_1^{2}+0.152b_1^{3})\, ,\\
	b_{\G_{2}} &= \bar{b}_{\G_{2}}(0.423-1.000b_1+0.310b_1^{2}+0.003b_1^{3})\, ,\\
	b_{\G_{2}\delta} &= \bar{b}_{\G_{2}\delta}(-0.344+1.333b_1-0.531b_1^{2}-0.005b_1^{3})\, ,\\
	b_{\G_{3}} &= \bar{b}_{\G_{3}}(0.349-0.349b_1)\, ,\\
	b_{\Gamma_{3}} &= \bar{b}_{\Gamma_{3}}(-0.535+0.535b_1)\, .
\end{align}
We show the redshift evolution of these bias parameters in Fig.~\ref{fig:bias}.
As mentioned in the main text, the biases $b_{\delta^3}$, $b_{\G_2\delta}$ and $b_{\G_3}$ lead to shape contributions that are degenerate with the other shapes at one loop. 

\section{Loop Integrals}\label{app:loop}

In this appendix, we collect the expressions for one-loop power spectra in the presence of massive neutrinos. A fully systematic treatment for massive neutrinos has been developed in~\cite{Blas:2014hya, Senatore:2017hyk}, This makes use of scale-dependent Green's functions, and the computational details are rather involved. Instead, we make a few assumptions to simplify numerical evaluation of the integrals. Namely, we use the EdS approximation that allows us to use the standard SPT kernels, but use the exact, scale-dependent linear power spectrum in the integrands. Moreover, we pull the scale-dependent growth functions out of the integrals, so that the redshift dependence factorizes. 

With the assumption of the factorized growth function, all one-loop integrals scale as 
\begin{equation}
	P_{1\text{-loop}}^{\O\O'}(z,z',k) = D_+^2(z,k)D_+^2(z',k)P_{1\text{-loop}}^{\O\O'}(k)\, .
\end{equation}
The purely momentum-dependent part of the one-loop integrals then take the standard form~\cite{Assassi:2014fva}
\begin{align}
	P_{22}^{\O\O'}\!(k) &= \textstyle\int_\q F_2^2(\q,\k-\q)P_{11}^{\O\O'}\!(q)P_{11}^{\O\O'}\!(|\k-\q|)\, ,\\
	P_{13}^{\O\O'}\!(k) &= 3P_{11}^{\O\O'}\!(k)\textstyle\int_\q F_3(\q,-\q,\k)(P_{11}^{\O\O}(q)+P_{11}^{\O'\O'}\!(q))\, ,\\
	{\cal I}_{\delta^2}^{\O\O'}\!(k) &=2\textstyle\int_\q F_2(\q,\k-\q)P_{11}^{\O\O'}\!(q)P_{11}^{\O\O'}\!(|\k-\q|)\, ,\\
	{\cal I}_{\G_2}^{\O\O'}\!(k) &= 2\textstyle\int_\q L_2(\q,\k-\q)F_2(\q,\k-\q)P_{11}^{\O\O'}\!(q)P_{11}^{\O\O'}\!(|\k-\q|)\, , \\
	{\cal F}_{\G_2}^{\O\O'}\!(k) &=4P_{11}^{\O\O'}\!(k)\textstyle\int_\q L_2(\q,\k-\q)F_2(\k,-\q)P_{11}^{\O\O'}\!(q)\, ,\\
	{\cal I}_{\delta^2\delta^2}^{\O\O'}\!(k) &= 2\textstyle\int_\q P_{11}^{\O\O'}\!(q)P_{11}^{\O\O'}\!(|\k-\q|)\, ,\\
	{\cal I}_{\G_2\G_2}^{\O\O'}\!(k) &=2\textstyle\int_\q L_2(\q,\k-\q)^2 P_{11}^{\O\O'}\!(q)P_{11}^{\O\O'}\!(|\k-\q|)\, ,\\
	{\cal I}_{\delta^2\G_2}^{\O\O'}\!(k) &= 2\textstyle\int_\q L_2(\q,\k-\q)P_{11}^{\O\O'}\!(q)P_{11}^{\O\O'}\!(|\k-\q|)\, .
\end{align}
Note that the integral $P_{13}^{\O\O'}$ that involves perturbing $\delta$ to cubic order contains the average between the two power spectra $P_{11}^{\O\O}$ and $P_{11}^{\O'\O'}$ in the loop integrand. These integrals can be computed efficiently with FFTLog~\cite{Simonovic:2017mhp}. This needs to be done only once, and the resulting functions provide inputs for the angular power spectra at one loop. 

\section{Detailed Parameter Forecasts}\label{app:more}
In this appendix, we provide further details of our forecast. In particular, we present and compare the forecasted constraints for three individual experiments: BOSS, DESI, and LSST.

\subsection{Experimental Specifications}\label{sec:exp}

This subsection summarizes the specifications of the CMB experiments and galaxy surveys that we consider in our analysis.

\subsubsection{Planck \& CMB-S4}\label{sec:CMB}

We consider two different CMB experiments: Planck~\cite{Akrami:2018vks} and CMB-S4~\cite{Abazajian:2016yjj,Akrami:2019izv}. We summarize their relevant experiment configurations in Tab.~\ref{tab:CMBparam}. For the CMB lensing, we use the publicly available code {\tt quicklens} to compute the reconstruction noise from the minimum variance quadratic estimator of \cite{Hu:2001kj, Okamoto:2003zw}. We also consider the improvement from the iterative lensing reconstruction \cite{Hirata:2003ka, Smith:2010gu} on $EB$ noise by rescaling. The left panel of Fig.~\ref{fig:windowLSS} shows the CMB lensing power spectrum together with the reconstruction noise. We see that Planck is noise dominated for the full multipole range, while CMB-S4 is signal dominated up to $\ell\sim 10^3$. These noise levels lead to signal-to-noise of approximately 40 and 400 for Planck and CMB-S4, respectively. The parameters $\ell_{\text{min}}$ and $f_{\text{sky}}$ for CMB lensing are set to be the same as those of the corresponding galaxy survey. 

\begin{table}[h!]
\centering
\begin{tabular}{c c c c c c c c c c c}
\toprule
  & $\theta_b$ [$'$] & $\Delta_T$ [$\mu$K$'$] & $\Delta_{E,B}$ [$\mu$K$'$] & $f_{\rm sky}$  & $\ell_{\rm min}$ & $\ell_{\rm max}^T$ & $\ell_{\rm max}^{E,B}$ \\
\midrule 
Planck & 5 & 43 & 81 & 0.65 & 2 &  2500 & 2500\\ 
S4 & 1 & 1 & 1.4 & 0.4 & 50& 3000 & 5000\\ 
\bottomrule
\end{tabular} 
\caption{Experimental specifications of Planck and CMB-S4.}
\label{tab:CMBparam}
\end{table} 

\subsubsection{BOSS}
For the {\it Baryon Oscillation Spectroscopic Survey} (BOSS), we use their spectroscopic sample of luminous red galaxies (LRG) and take the number density as given in~\cite{Font-Ribera:2013rwa} with the sky coverage of $9329\, \text{deg}^{2}$. The effective redshift range for the LRG sample is $[0, 0.8]$, which we split into 4 non-overlapping tomographic bins with edges given by the set $\{0, 0.2, 0.4, 0.6, 0.8\}$. The linear bias is assumed to evolve as $b_\delta(z)=1.7 D_+^{-1}(z)$. 	

\subsubsection{DESI}

For the {\it Dark Energy Spectroscopic Instrument} (DESI), we use their spectroscopic sample of emission line galaxies (ELG) and take the number density as given in Table 2.3 of \cite{Aghamousa:2016zmz} with the sky coverage of $14000\, \text{deg}^{2}$. 
The effective redshift range for the ELG sample is $[0.6, 1.7]$, which we split into 4 non-overlapping tomographic bins with edges given by the set $\{0.6, 0.9, 1.2, 1.5, 1.7\}$. The linear bias is assumed to evolve as $b_\delta(z)=0.84 D_+^{-1}(z)$.

\subsubsection{Vera Rubin Observatory}\label{sec:LSS}

For LSST, we use the so-called ``Gold'' sample of galaxies~\cite{Abell:2009aa}, whose redshift distribution is modeled by
\begin{equation}
\frac{\d n}{\d z} \propto \frac{1}{2z_{0}}\left(\frac{z}{z_{0}}\right)^{2}e^{-z/z_{0}}\, ,\label{Gold}
\end{equation}
with $z_{0}=0.3$. This gives $\bar n = \int \frac{\d n}{\d z}\,\d z=40\, \text{arcmin}^{-2}$ for the effective redshift range of $[0,7]$ and the sky coverage of $18000\, \text{deg}^{2}$. We split this into 16 non-overlapping tomographic bins with edges given by the set $\{0, 0.2, 0.4, 0.6, 0.8, 1, 1.2, 1.4, 1.6, 1.8, 2, 2.3, 2.6, 3, 3.5, 4, 7\}$. We employ a simple evolution model for the linear bias, with $b_{1}(z)=\bar b_{1}(1+z)$.

\begin{figure}[t!]
    \centering
            \includegraphics[scale=.74]{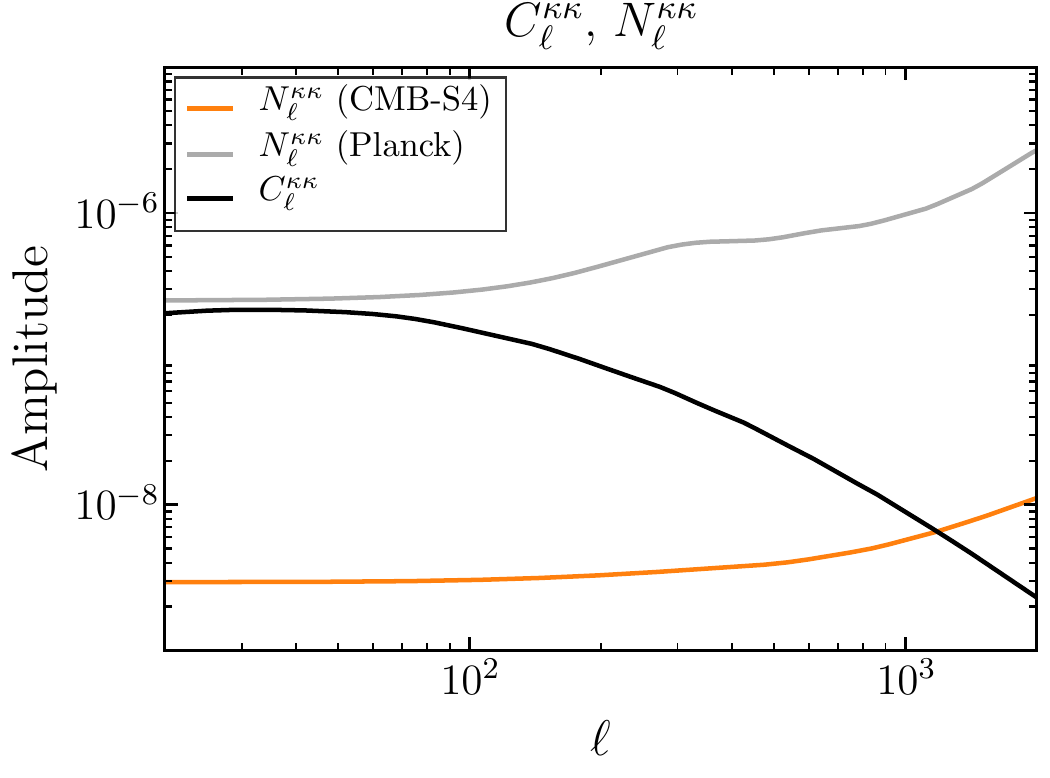}         
                  \includegraphics[scale=.74]{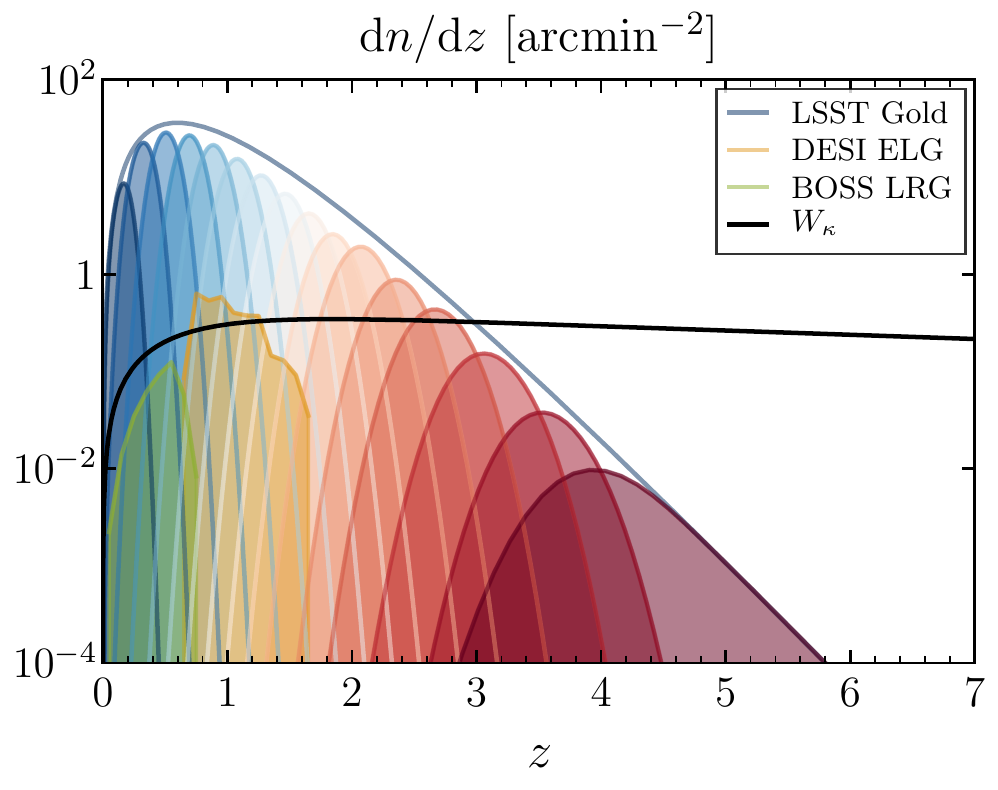}   
         \\
    \caption{{\it Left panel:} Comparison between the signal $C_{\ell}^{\kappa\kappa}$ with the CMB lensing reconstruction noise $N_{\ell}^{\kappa\kappa}$ for Planck and CMB-S4 from the minimum variance quadratic estimator of~\cite{Hu:2001kj, Okamoto:2003zw}. {\it Right panel:} Galaxy number density for the different LSS experiments considered in this work and the window function for CMB lensing. The blue/red filled curves show the redshift distributions of LSST Gold galaxy samples, which lead to the galaxy power spectra shown in Fig.~\ref{fig:ClLSST}.}
    \label{fig:windowLSS}
\end{figure}

To account for photometric redshift errors, we follow the prescription of~\cite{Ma:2005rc} of convolving the window function with the probability distribution function $p(z_{\rm ph}|z)$ of the photometric redshift $z_{\rm ph}$ at a given $z$. The true redshift galaxy distribution for $i$-th photo-$z$ bin $[z^{(i)},z^{(i+1)}]$ then becomes
\begin{align}
	\frac{\d n_i}{\d z} = \frac{\d n}{\d z} \int_{z^{(i)}}^{z^{(i+1)}}\! \d z_{\rm ph}\, p(z_{\rm ph}|z)\, .
\end{align}
It is convenient to consider a Gaussian probability distribution function
\begin{align}
	p(z_{\rm ph}|z) =\frac{1}{\sqrt{2\pi}\sigma_z}\exp\left[-\frac{(z-z_{\rm ph}-z_{\rm bias})^2}{2\sigma_z^2}\right]\, ,
\end{align}
where $z_{\rm bias}$ is a bias parameter and $\sigma_z$ is the width, both of which can be redshift dependent. This then gives
\begin{align}
	\frac{\d n_i}{\d z} = \frac{1}{2}\frac{\d n}{\d z} \left[{\rm erf}\left(\frac{z-z^{(i)}-z_{\rm bias}}{\sqrt{2}\sigma_z}\right)-{\rm erf}\left(\frac{z-z^{(i+1)}-z_{\rm bias}}{\sqrt{2}\sigma_z}\right)\right]\, ,
\end{align}
with `${\rm erf}$' the error function. In our forecast, we set $z_{\rm b}=0$ and $\sigma_z=0.05(1+z)$. We show the corresponding window functions in the right panel of Fig.~\ref{fig:windowLSS}. The total effective number density after convolution is $\bar{n}\approx25$ $\text{arcmin}^{-2}$, which is $37.5\%$ smaller than the one without photometric errors.

\subsection{Results and Comparison}

\begin{table}[t!]
\centering
\begin{tabular}{c c c c c c c c c c}
\toprule
 & \multicolumn{3}{c}{LSS} & \multicolumn{3}{c}{$\times$ CMB Lensing} & \multicolumn{3}{c}{+ CMB T\&P} \\
 	\cmidrule(lr){2-4} \cmidrule(lr){5-7}\cmidrule(lr){8-10}
  & BOSS & DESI & LSST & BOSS & DESI & LSST  & BOSS & DESI & LSST\\
\midrule 
$\sigma(H_0)$ [km/s/Mpc] & 119 & 47 & 4.35 & 39 & 6.01 & 1.71 & 1.75 & 0.96 & 0.31\\
$10^5\sigma(\omega_b)$ & 12930 & 5325 & 498 & 4266 & 709 & 21 & 13 & 2.8 & 2.5\\
$10^4\sigma(\omega_c)$ & 4747 & 1873 & 190 & 1536 & 230 & 57 & 13 & 7 & 3.5\\
$10^{9}\sigma(A_s)$& 8.49 & 2.34 & 0.54 & 1.42 & 0.24 & 0.030 & 0.020 & 0.066 & 0.005\\
$10^4\sigma(n_s)$ & 10967 & 4314 & 537 & 2728 & 301 & 73 & 34 & 22 & 10\\
\midrule
$\sigma(M_{\nu})$ [meV] & 8018 & 1984 & 283 & 1693 & 371 & 102 & 159 & 96 & 21\\
$\sigma(f_{\text{NL}})$ & 278 & 75 & 3.46 & 140 & 35 & 2.01 & 108 & 29 & 1.85\\
\bottomrule
\end{tabular} 
\caption{Comparison of parameter constraints from tree-level power spectra between different experiments. For CMB T\&P, we use Planck for BOSS and CMB-S4 for DESI and LSST.} 
\label{tab:param3}
\end{table} 

In Table \ref{tab:param3}, we display a comparison between our forecast results for BOSS, DESI, and LSST. For the purpose of making a fair comparison, here we have only used the tree power spectrum information for all of the experiments.
The three galaxy surveys all exhibit a similar trend when the CMB information has been added: the parameter constraints are significantly improved compared to using galaxy-only statistics. 
Looking at the combined LSS and CMB constraints, we see that LSST will be capable of substantially improving parameter constraints over BOSS and DESI (for the samples considered here). 
This can be understood from the fact that LSST will have access to a much higher number density of galaxies (see Fig.~\ref{fig:windowLSS}), which leads to much reduced shot noise in each tomographic bin.

\section{Analytical Method}\label{app:analytic}

The main advantage of the FFTLog method is that it allows the momentum integrals to be done analytically. When dealing with the scale-dependent growth function due to massive neutrinos, it is useful to combine FFTLog with the polynomial approximation introduced in \S\ref{sec:poly}, which allows us to evaluate the remaining integrals with fully analytic formulas. 
In this appendix, we provide details of this method for angular power spectra in \S\ref{app:PS} and comment on its computational efficiency in \S\ref{app:perf}.

\subsection{Angular Power Spectra}\label{app:PS}

\begin{figure}[t!]
    \centering
         \includegraphics[width=6.5cm]{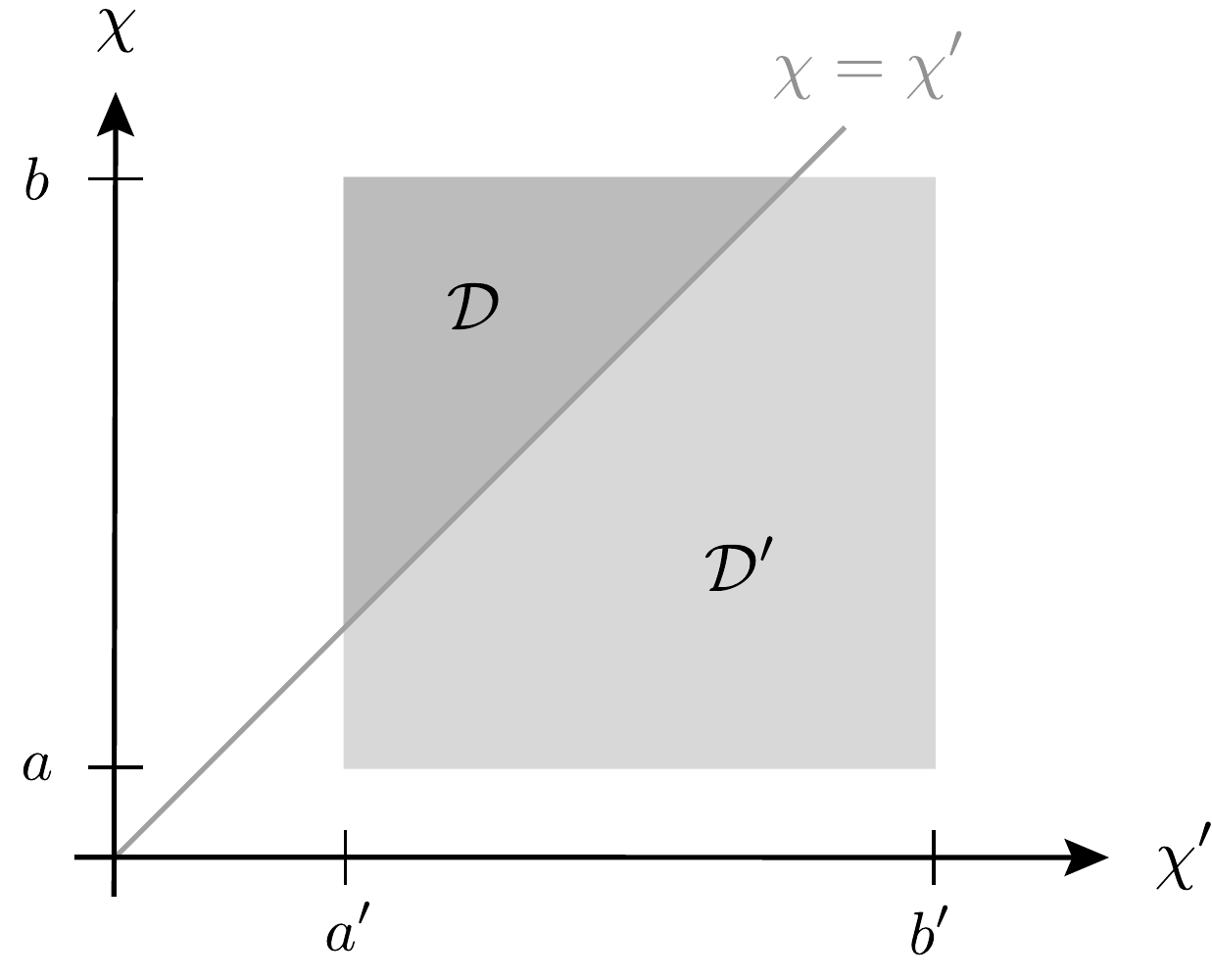}
    \caption{Domains of integration $\D$ and $\D'$ for the line-of-sight integral \eqref{Kell} in the $(\chi',\chi)$-plane. By assumption, $\D'$ has a larger area than $\D$. The limit in which the $\D$ and $\D'$ become triangular and have an equal area corresponds to auto-spectra. The opposite limit in which $\D$ vanishes corresponds to cross-spectra with non-overlapping bins. A general configuration describes cross-spectra with overlapping bins.}
    \label{fig:domain}
\end{figure}

Recall that the angular power spectrum at tree level takes the form (c.f.~\eqref{Cell}; dropping operator labels to avoid clutter)
\begin{align}
	C_\ell &= \frac{1}{2\pi^2}\int_0^\infty\d\chi\, W(\chi)\int_0^\infty\d\chi'\, W(\chi') I_\ell(\chi,\chi';0)\, ,\\ 
	I_{\ell}(\chi,\chi';n)&\equiv  4\pi\int_0^\infty\d k\, k^{2+n} j_\ell(k\chi)j_{\ell}(k\chi') P(\chi,\chi',k)\,.
\end{align}
To simplify the integrals, we first write $P(\chi,\chi',k)=D_+(\chi,k)D_+(\chi',k)P(k)$ and then apply the polynomial approximation to the window function together with the growth function as
\begin{align}
	D_+(\chi,k)W(\chi) \simeq \sum_{p=0}^{N_{\rm poly}} w_p(k) \chi^p\, .\label{Wpoly}
\end{align} 
We then apply the FFTLog decomposition to the whole $k$-dependent part as
\begin{align}
	w_p(k)w_q(k)P(k) \simeq \sum_{n=-N_\eta}^{N_\eta} c_{npq}\, k^{-b+\eta_{npq}}\, ,
\end{align}
for each $p,q$, with constant coefficients $c_{npq}$. The number of independent FFTLog transforms that need to be taken is $(N_{\rm poly}+1)(N_{\rm poly}+2)/2$. The cross power spectrum between the tomographic bins $i$ and $j$ can then be computed as
\begin{align}
		C_\ell &\simeq \frac{1}{2\pi^2} \sum_{n,p,q}c_{npq}\int_{\mathbb{Z}_i}  \d \chi\,\chi^{p}\int_{\mathbb{Z}_j} \d \chi'\, \chi'^{q-\nu_{npq}} \,\I_\ell(\nu_{npq},\tfrac{\chi}{\chi'})\, ,\label{Cell2}
\end{align}
where $\I_\ell$ was defined in~\eqref{WSint}, $\nu_{npq}\equiv 3-b+i\eta_{npq}$, and $\int_{\mathbb{Z}_i}$ denotes the integral over the $i$-th tomographic bin. 
We see that the $\chi,\chi'$ integrals are of the advertised form in~\eqref{eq:poly}. 
The task of computing $C_\ell$ then boils down to evaluating integrals of the form
\begin{align}
	\K_\ell  &\equiv \int^{b}_{a}\d \chi\, \chi^{\alpha-1} \int^{b'}_{a'} \d \chi'\, \chi'^{\beta-1}\, \I_\ell (\nu,\tfrac{ \chi}{\chi'})\, ,\label{Kell}
\end{align}
where we have suppressed the arguments on the left-hand side for brevity. Without loss of generality, suppose that $a\le a'$ and $b\le b'$. As depicted in Fig.~\ref{fig:domain}, the domain of integration $[a,b]\times [a',b']$ is in general divided into two subregions which we label by $\D$ and $\D'$, with the line $\chi=\chi'$ separating them. 
Note that the argument of $\I_\ell$ in \eqref{Kell} becomes greater than one in $\D$, in which case we need to use the inversion formula $\I_\ell(\nu,w)=w^{-\nu}\I_\ell(\nu,\tfrac{1}{w})$ to appropriately split the integrals.

Our task is to derive an analytic expression for the line-of-sight integral \eqref{Kell}. This integral falls into two types: ones with vanishing and non-vanishing $\D$. The former correspond to cross-spectra with non-overlapping tomographic bins, while the latter encapsulates those with overlapping tomographic bins as well as auto-spectra. 
The basic building blocks for writing the result in terms of analytic expressions are the following indefinite integrals:
\begin{align}
 \int\d \chi\, \chi^{\alpha-1}\! \int \d \chi' \chi'^{\beta-1} \I_\ell (\nu,\tfrac{ \chi}{\chi'}) \,=\, &\begin{cases} \sX_\ell(\chi,\chi') & \chi/\chi' \le 1\, , \\[3pt] \widetilde\sX_\ell(\chi,\chi')\equiv \sX_\ell(\chi',\chi)|^{\alpha\to \beta+\nu}_{\beta\to \alpha-\nu}\phantom{\hskip 0pt} & \chi/\chi' > 1 \, , \end{cases}\\[2pt]
	\int\d \chi\, \chi^{\alpha-1} \left[\int \d \chi' \chi'^{\beta-1} \I_\ell (\nu,\tfrac{ \chi}{\chi'})\right]_{\!\chi'=\chi}  =\, &\begin{cases} \sY_\ell(\chi) & \chi/\chi' \le 1\, , \\[3pt] \widetilde\sY_\ell(\chi)\equiv -\sY_\ell(\chi)|^{\alpha\to \alpha+2\beta+\nu}_{\beta\to -\beta-\nu}\phantom{\hskip 4pt} & \chi/\chi' > 1 \, , \end{cases}
 \end{align}
 where we have defined the functions
 \begin{align}
 &\sX_\ell(\chi,\chi')  \equiv \frac{\chi^\alpha \chi'^\beta}{\alpha+\beta} \big[F_\ell(\beta,\nu,\tfrac{\chi}{\chi'}) - (\beta\leftrightarrow{-\alpha}) \big]\,, \quad  	\sY_\ell(\chi) \equiv \frac{\chi^{\alpha+\beta}}{\alpha+\beta}\hs F_\ell(\beta,\nu,1)\, ,\label{XY} \\[5pt]
	&  \text{with} \qquad  F_\ell(\beta,\nu,w)\equiv \frac{2^{\nu-1}\pi^2\Gamma(\ell+\frac{\nu}{2})}{\Gamma(\frac{3}{2}+\ell)\Gamma(\frac{3}{2}-\frac{\nu}{2})}\frac{w^\ell}{\beta-\ell}\, {}_3F_2\Bigg[\begin{array}{c} \frac{\ell-\beta}{2},\hs \frac{\nu-1}{2},\hs \ell+\frac{\nu}{2} \\[2pt]  \frac{\ell-\beta}{2}+1,\hs \ell+\frac{3}{2}\end{array}\Bigg|\, w^2\Bigg]\, ,
 \end{align}
and the generalized hypergeometric function ${}_3F_2$. The integral \eqref{Kell} can then be written as
\begin{align}\label{Kell2}
	\K_\ell &= \sX_\ell(a,a')-\sX_\ell(a,b')-\sX_\ell(a',a')+\sX_\ell(a',b')+ \sY_\ell(a')-\sY_\ell(b)-\sX_\ell(a',b')+\sX_\ell(b,b')\nn[3pt]
	&\quad +\wt\sX_\ell(a',a')-\wt\sX_\ell(b,a')-\wt\sY_\ell(a')+\wt\sY_\ell(b)\, ,
\end{align}
where the first and second lines calculate the integrals for the domains $\D'$ and $\D$, respectively. For cross-spectra with non-overlapping bins, only the first four terms in \eqref{Kell2} contribute, whereas only the last eight terms survive for auto-spectra.

\subsection{Performance and Precision}\label{app:perf}

\begin{figure}[t!]
    \centering
         \includegraphics[scale=.8]{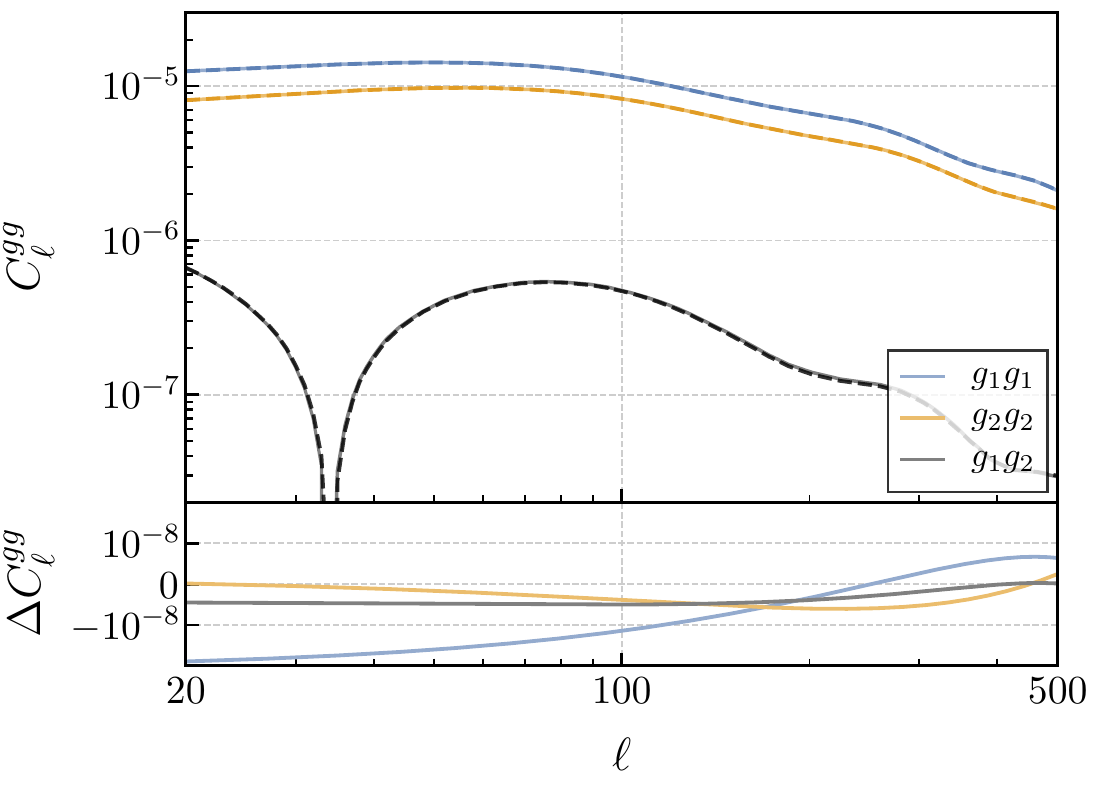}
    \caption{Angular galaxy power spectra at tree level with the line-of-sight integrals evaluated using the analytic (solid lines) and numerical methods (dashed lines). We use the LSST non-overlapping window functions over the redshift intervals $[1.8,2.0]$ and $[2.0,2.3]$ for galaxy overdensities denoted by $g_1$ and $g_2$, respectively. The bottom panel shows the absolute differences between the computed power spectra. The parameters used for numerical computation are specified in Tab.~\ref{tab:performance}.}
    \label{fig:AnalyticvsGaussQuad}
\end{figure}

Let us compare the two methods for computing the angular power spectrum \eqref{Cell2}, which involve: (i) numerically integrating the line-of-sight integrals using Gaussian quadrature, which we dub ``numerical'', and (ii) using the analytic expression given in~\eqref{Kell2}, which we dub ``analytical''.
A comparison of the angular power spectrum computed with the numerical and analytic methods is shown in Fig.~\ref{fig:AnalyticvsGaussQuad}, for two representative tomographic bins of LSST. As we can see, the agreement is excellent. 

Some parameters and benchmark performance of these methods are presented in Tab.~\ref{tab:performance}. In particular, we see that the use of the analytic formula \eqref{Kell2} can be roughly 15 times computationally more efficient than the numerical integration method when dealing with the scale-dependent growth function due to massive neutrinos. 
The line-of-sight integrals in \eqref{Cell2} are two dimensional, so the computational cost scales as $N_{\chi}^2 N_{{}_2F_1}$ for the numerical method, where $N_\chi$ and $N_{{}_pF_q}$ are the numbers of terms required for the numerical $\chi$-integral and the hypergeometric series ${}_pF_q$ for convergence, respectively.  
This is to be compared with the scaling that goes as $N_{{}_3F_2}$ for the analytical method. The hypergeometric ${}_3F_2$ converges very fast for the multipole range considered, with $N_{{}_3F_2}\ll N_{\chi}^2 N_{{}_2F_1}$.

\begin{table}[t!]
\begin{center}
\begin{tabular}{c c c c c c c}
\toprule
\hskip 5pt  Method\phantom{\hskip 5pt} &\phantom{\hskip 2.5pt} $M_\nu$ \phantom{\hskip 2.5pt} &  $N_\eta$ & $N_\chi$ & $N_{\rm poly}$ &  $N_\ell$ & \phantom{\hskip 5pt} Time \phantom{\hskip 5pt} \ \\[3pt]
\midrule
Numerical & $-$ & 100 & 80 & $-$ & 30 & $1\,$min \\[3pt]
Analytical & $-$  & 100 & $-$ & 3  & 30 &  $10\,$s\\[3pt]
\midrule
Numerical & \checkmark & 100 & 80 & 3 & 30 &  $10\,$min \\[3pt]
Analytical & \checkmark & 100 & $-$ & 3 & 30 & $10\,$s \\[3pt]
\bottomrule
\end{tabular}
\end{center}
\vspace{-0.5cm}
	\caption{Parameters and benchmark performance for different computational methods. The last column denotes the approximate time taken for evaluating the galaxy angular power spectrum $C_{\ell}^{gg}$ at $N_\ell$ sampled points in the range $20\leq\ell\leq500$ using our \texttt{Mathematica} code on a laptop, with pre-computed FFTLog coefficients.}
	\label{tab:performance}
\end{table}

While the analytical method generally performs superiorly to the numerical method, let us also mention some limitations of the former approach.
Due to the polynomial approximation of the window function, the computational cost scales as $N_{\rm poly}^2$. For windows such as Gaussian and the photometric window functions that decay fast near the boundaries of tomographic bins, one needs to use a relatively higher degree of the polynomial for convergence, with $N_{\rm poly}\simeq 15$. 
Secondly, even though the generalized hypergeometric function ${}_3F_2[\cdots|w]$ is absolutely convergent at $w=1$ for the case at hand, its convergence rate becomes rather slow for $\ell\lesssim 10$. 
We have not attempted to exhaust all possible hypergeometric identities in this work, and further algebraic manipulations of  ${}_3F_2$ may help optimize the calculation at very low $\ell$.

\newpage
\bibliographystyle{utphys}
\bibliography{CMBxLSS}

\end{document}